\documentstyle[preprint,tighten,eqsecnum,aps,floats,psfig,epsfig,amssymb]{revtex}

\begin{document}
\draft
\title{
Multicritical behavior in frustrated spin systems with noncollinear order}
\author{Pasquale Calabrese,$^{1}$ 
Andrea Pelissetto,$^2$ and Ettore Vicari$\,^3$ }
\address{$^1$ 
R.~Peierls Center for Theoretical Physics, University of Oxford, \\
1 Keble Road, Oxford OX1 3NP, United Kingdom.}
\address{$^2$ Dip. Fisica dell'Universit\`a di Roma ``La Sapienza"  and INFN, \\
P.le Moro 2, I-00185 Roma, Italy}
\address{$^3$
Dip. Fisica dell'Universit\`a di Pisa
and INFN, V. Buonarroti 2, I-56127 Pisa, Italy}
\address{
\bf e-mail: \rm 
{\tt calabres@df.unipi.it,}
{\tt Andrea.Pelissetto@roma1.infn.it},
{\tt vicari@df.unipi.it}
}

\date{\today}

\maketitle

\begin{abstract}
We investigate the phase diagram and, in particular, the nature of the 
the multicritical point in three-dimensional frustrated
$N$-component spin models with noncollinear order in the presence of
an external field, for instance easy-axis stacked triangular
antiferromagnets in the presence of a magnetic field along the easy
axis.  For this purpose we study the renormalization-group flow in a
Landau-Ginzburg-Wilson $\phi^4$ theory with symmetry
O(2)$\otimes$[${\mathbb Z}_2 \oplus$O($N-1$)] that is expected to
describe the multicritical behavior.  We compute its $\overline{\rm
MS}$ $\beta$ functions to five loops.  For $N\ge 4$, their analysis
does not support the hypothesis of an effective enlargement of the
symmetry at the multicritical point, from O(2)$\otimes$[${\mathbb Z}_2
\oplus$O($N-1$)] to O(2)$\otimes$O($N$).  For the physically
interesting case $N=3$, the analysis does not allow us to exclude the
corresponding symmetry enlargement controlled by the O(2)$\otimes$O(3)
fixed point. Moreover, it does not provide evidence for any other
stable fixed point.  Thus, on the basis of our field-theoretical
results, the transition at the multicritical point is expected to be
either continuous and controlled by the O(2)$\otimes$O(3) fixed point
or to be of first order.

\end{abstract}



\section{Introduction}
\label{intro}

\begin{figure}
\centerline{\epsfig{file=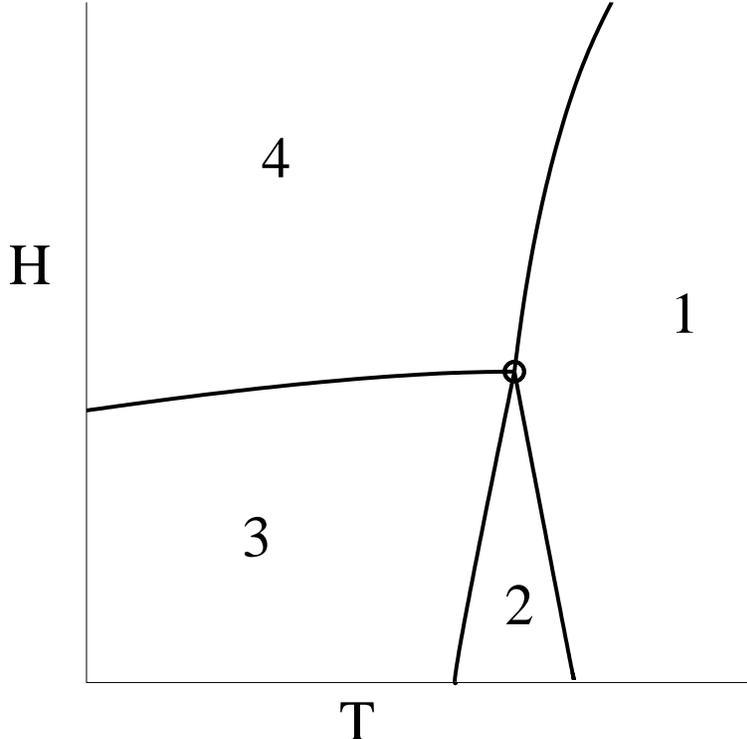,width=10truecm}}
\caption{The experimentally observed phase diagram in easy-axis 
STAs as a function of temperature $T$ and magnetic field $H$.}
\label{figmcp}
\end{figure}

Frustrated spin models with noncollinear order, such as easy-axis
stacked triangular antiferromagnets (STAs), reveal a quite complex
phase diagram in the presence of an external magnetic field $H$ along
the easy axis, see, e.g.,
Refs.~\cite{CP-97,Kawamura-98,JRF-79,BWL-93,EFS-94,ESMK-97,PCP-90,%
KGA-93,AAMKG-94}.  A model of these systems is obtained by considering
a stacked triangular lattice, three-component spins ${s}_i$ defined at
the sites of the lattice satisfying ${s}_i\cdot {s}_i = 1$, and the
Hamiltonian
\begin{equation}
{\cal H}_{\rm STA} = {\beta\over2} \sum_{ij} J_{ij} {s}_i \cdot {s}_j 
    + \sum_i [D s_{i,z}^2 + H s_{i,z}],
\label{lattmodel}
\end{equation}
with an antiferromagnetic hopping term $J_{ij}$.  For small magnetic
fields, one observes two critical lines that are expected to belong to
the XY universality class according to the theoretical analysis
\cite{PHC-88,PC-90,KCP-90,MPC-93}.  For large magnetic fields, there
is instead a single critical line that is expected to belong to the
O(2)$\otimes$O(2) universality class, which is characterized by the
symmetry-breaking pattern O(2)$\otimes$O(2) $\rightarrow$ O(2)$_{\rm
diag}$.  Finally, the large-$H$ and small-$H$ domain are separated by
a first-order spin-flop line. These four critical lines meet at a
tetracritical point, see Fig.~\ref{figmcp}.  The authors of
Ref.\cite{KCP-90} argued that the critical behavior at the
multicritical point belongs to the O(2)$\otimes$O(3) universality
class, with symmetry-breaking pattern O(2)$\otimes$O(3) $\rightarrow
{\mathbb Z}_2$$\otimes$O(2)$_{\rm diag}$, described by the Hamiltonian
\cite{Kawamura-88}
\begin{eqnarray}
{\cal H}_{\rm sym} =&& \int d^d x \left\{ 
  {1\over2}
\sum_{ai} \left[ \sum_\mu (\partial_\mu \Phi_{ai})^2 + r \Phi_{ai}^2 
      \right]   
\right.
\nonumber \\
&& \left.
+ {g_{1,0}\over 4!}  ( \sum_{ai} \Phi_{ai}^2)^2 
+ {g_{2,0}\over 4!}  \left[ \sum_{i,j} 
( \sum_a \Phi_{ai} \Phi_{aj})^2 - (\sum_{ai} \Phi_{ai}^2)^2 \right]
\right\} ,
\label{LGWH-chirale}
\end{eqnarray}
where $\Phi_{ai}$ is a $3\times2$ matrix, i.e., $a=1,2,3$ and $i=1,2$.
The analysis of Ref.~\cite{KCP-90} is however not complete, since only
the quadratic perturbations of the O(2)$\otimes$O(3) fixed point (FP)
were considered.  In this paper we reconsider the issue, by performing
a complete analysis of all quadratic and quartic perturbations induced
by the easy-axis anisotropy, including also 
those terms that are absent in the theoretical analysis of Ref.~\cite{KCP-90}.

The Landau-Ginzburg-Wilson (LGW) Hamiltonian that describes the
multicritical behavior is the most general Hamiltonian with symmetry
O(2)$\otimes$[${\mathbb Z}_2$$\oplus$O(2)].  It is given by
\begin{eqnarray}
{\cal H} &=& \int d^d x \left\{ 
  {1\over2}
\sum_{ai} \left[ \sum_\mu (\partial_\mu \phi_{ai})^2 + r_\phi \phi_{ai}^2 
      \right]+
{1\over2}
\sum_i\left[ \sum_\mu (\partial_\mu \psi_i)^2 + r_\psi \psi_i^2 \right]
\nonumber   \right. \\&&
+ {u_0\over 4!}  ( \sum_{ai} \phi_{ai}^2+\sum_i \psi_i^2)^2 
+ {v_0\over 4!}  \left[ \sum_{i,j} 
( \sum_a \phi_{ai} \phi_{aj} +\psi_i\psi_j)^2 - 
(\sum_{ai} \phi_{ai}^2 + \sum_i \psi^2_i)^2 \right]
\nonumber\\&&
\left. 
+{w_0\over4!} ( \sum_{ai} \phi_{ai}^2)^2 
+{y_0\over4!} (  \sum_i \psi_i^2 )^2 
+{z_0\over4!} \sum_{ij} [(\sum_a \phi_{ai} \phi_{aj})^2 - \sum_a \phi_{ai}^2
\sum_a \phi_{aj}^2 ]
\right\} ,
\label{LGWH}
\end{eqnarray}
where $\phi_{ai}$ and $\psi_i$ are real fields with $a=1,2$ and
$i=1,2$.  The LGW Hamiltonian (\ref{LGWH}) can also be obtained from
model (\ref{lattmodel}) by performing a Hubbard-Stratonovitch
transformation \cite{Baker_62,Hubbard_72} and an expansion in terms of
the critical modes.

The critical behavior at the multicritical point is determined by the
stable FP of the renormalization-group (RG) flow when both $r_\phi$
and $r_\psi$ are tuned to their critical value.  See,
e.g., Refs.~\cite{KNF-76,CPV-03} and references therein for a discussion
of the field-theoretical (FT) approach to multicritical phenomena. If no
stable FP exists or if the system is not in the attraction domain of
the stable FP, the transition at the multicritical point
is expected to be of first order. The
hypothesis of the effective enlargement of the symmetry
\begin{equation}
{\rm O}(2)\otimes[{\mathbb Z}_2 \oplus {\rm O}(2)] 
\;
\rightarrow 
\;
{\rm O}(2)\otimes{\rm O}(3)
\label{symenl}
\end{equation}
at the multicritical point
requires that the O(2)$\otimes$O(3) chiral FP is stable with respect
to the quartic terms that break O(2)$\otimes$O(3) to
O(2)$\otimes$[${\mathbb Z}_2\oplus$O(2)]---those proportional to
$w_0$, $y_0$, and $z_0$ in Hamiltonian (\ref{LGWH}).  If this does not
occur, the O(2)$\otimes$O(3) FP does not control the multicritical
behavior for generic values of the Hamiltonian parameters.  As a
consequence, the effective enlargement of the symmetry to O(2)$\otimes$O(3) at
the multicritical point requires an additional tuning of the
parameters: beside tuning $r_\phi$ and $r_\psi$, at least one more
Hamiltonian parameter must be properly fixed to decouple the
additional relevant interaction.

In this paper we investigate this issue by FT
methods.  We consider the more general theory in which the order
parameter $\phi_{ai}$ is an $(N-1)\times 2$ matrix, i.e., $a=1,...N-1$
and $i = 1,2$, for $N\ge 3$.  In this case, setting $w_0=y_0=z_0=0$
and $r_\phi = r_\psi$, one recovers the O(2)$\otimes$O($N$)-symmetric
LGW Hamiltonian (\ref{LGWH-chirale}).  This theory has a stable FP
with attraction domain in the region $g_{2,0}>0$ describing a critical
behavior with symmetry-breaking pattern O(2)$\otimes$O($N$)
$\rightarrow$ O(2)$\otimes$O($N-2$)
\cite{Kawamura-88,PRV-01,CPPV-04}. We should mention that the
existence of this FP has been a controversial issue for quite a long
time; the different scenarios are reviewed in
Refs.~\cite{Kawamura-98,CPPV-04,TDM-00,PV-rev,DMT-03}.

In order to determine the RG flow of the
theory (\ref{LGWH}) in three dimensions and determine its
multicritical behavior, we consider
the minimal-subtraction ($\overline{\rm MS}$) scheme without
$\epsilon$ expansion (henceforth indicated as
$3d$-$\overline{\rm MS}$ scheme) in which no $\epsilon$ expansion is
performed and $\epsilon$ is set to the physical value $\epsilon=1$
\cite{SD-89}.  We use a symbolic manipulation program that generates
the diagrams and computes symmetry and group factors, and the
compilation of Feynman integrals of Ref.~\cite{KS-01}. This allows us
to compute the $3d$-${\overline{\rm MS}}$ $\beta$-functions to five
loops in the full model. The perturbative series are used to determine
the FP structure and, in particular, to investigate the existence of
stable FPs that may describe the critical behavior at the
multicritical point.  The $3d$-$\overline{\rm MS}$ scheme is
particularly convenient for the three-dimensional FT study of the
multicritical behavior. Indeed, the multicritical theory is simply
obtained by setting $r_\phi=r_\psi=0$, i.e. by considering the
massless theory.  Note that this is not correct in the
three-dimensional massive zero-momentum (MZM) scheme
\cite{footnoteMZM}, in which a proper tuning of $r_\phi$ and $r_\psi$
is needed.

We summarize the main results of this paper.  In order to check the
hypothesis of the effective enlargement of the symmetry
\begin{equation}
{\rm O}(2)\otimes[{\mathbb Z}_2 \oplus {\rm O}(N-1)] 
\;
\rightarrow 
\;
{\rm O}(2)\otimes{\rm O}(N)
\label{symenln}
\end{equation}
for generic $N$-component systems at the multicritical point, 
we study the stability properties
of the O(2)$\otimes$O($N$) chiral FP with respect to all quadratic and
quartic perturbations that are symmetric under the reduced symmetry
O(2)$\otimes$[${\mathbb Z}_2 \oplus$O($N-1$)].  The analysis of the
corresponding five-loop series does not support the stability of the
O(2)$\otimes$O($N$) chiral FP for any value $N\ge 4$, with increasing
confidence as $N$ increases.  For $N=3$ the results are not
conclusive. Our FT results do not allow us to
establish the stability properties of the O(2)$\otimes$O(3) FP, which
may be either stable or unstable.  In the former case, the
multicritical behavior would be controlled by the O(2)$\otimes$O(3) FP
if the transition is continuous.  In the latter case, we note that the
crossover exponent should be very small, $\phi_{4,4}\lesssim
0.1$. Therefore, if the effective quartic Hamiltonian parameters that
break the O(2)$\otimes$O(3) symmetry are small, the crossover from the
preasymptotic O(2)$\otimes$O(3) critical behavior to the eventual
asymptotic behavior is expected to be very slow, and one may observe
an effective O(2)$\otimes$O(3) critical behavior for a wide range of
reduced-temperature values.  We also perform a general Pad\'e-Borel
analysis of the RG flow for $N=3$, to investigate the existence of
stable FPs for generic values of the quartic couplings.  No evidence
of additional stable FPs is obtained.  Therefore, according to our FT
results, for $N=3$ the multicritical transition is either controlled
by the O(2)$\otimes$O(3) FP or is of first order.

The paper is organized as follows. In Sec.~\ref{sec2} we show how
the general Hamiltonian (\ref{LGWH}), that has been
written down on the basis of symmetry considerations, can be recovered 
from Hamiltonian (\ref{lattmodel}) 
of easy-axis STAs in a magnetic field along the
easy axis.  Then, we discuss the mean-field phase diagram, showing
that, beside the tetracritical behavior that was predicted in
Ref.~\cite{PHC-88}, the model also admits a bicritical and a
pentacritical phase diagram.  In Sec.~\ref{sec3} we focus on some
particular cases.  In Sec.~\ref{sec3.2} we discuss the stability of
the O(2)$\otimes$O($N$) FPs, and, in Sec.~\ref{sec3.3}, the stability
of the decoupled [O(2)$\otimes$O($N-1$)]$\oplus$O(2) FPs.  In
Sec.~\ref{sec4} we discuss the full model: in Sec.~\ref{sec4.1} we
consider the one-loop $\epsilon$ expansion, while in Sec.~\ref{sec4.2}
we numerically investigate the RG flow for $N=3$.  In Sec.~\ref{sec5}
we present our conclusions and critically discuss the experimental
results in view of our findings.  In App.~\ref{AppA} we report a
discussion of the mean-field diagram, in App.~\ref{AppB} we classify
all quadratic and quartic perturbations of the O(2)$\otimes$O($N$)
FPs that are O(2)-invariant.  Finally, in App.~\ref{AppC} we compute the RG
dimensions of all quadratic perturbations of the O(2)$\otimes$O($N$)
symmetric theory.

\section{Derivation of the general Hamiltonian and mean-field analysis} 
\label{sec2}

In this Section we derive the effective LGW Hamiltonian (\ref{LGWH})
for easy-axis STAs in a magnetic field along the easy axis described
by Hamiltonian (\ref{lattmodel}).  It can be obtained as usual by
first performing a Hubbard-Stratonovitch transformation
\cite{Baker_62,Hubbard_72}.  If $\Phi_a$, $a=1,2,3$, is an
unconstrained three-component field, the partition function can be
rewritten as
\begin{equation}
Z \sim \int \prod_a d\Phi_a\, 
    \exp\left[ {1\over 2 \beta} \sum_{ij} (J^{-1})_{ij}
    \Phi_i\cdot \Phi_j - \sum_i W(\Phi_i)\right],
\end{equation}
where 
\begin{equation}
W(\Phi) = - \ln \left[
   {\int d^3{s}\, \delta(s^2 - 1)\, \exp(\Phi\cdot s - D s^2_z - H s_z) \over 
    \int d^3{s}\, \delta(s^2 - 1)\, \exp(- D s^2_z - H s_z) } \right]\; .
\end{equation}
As usual, we now expand $W(\Phi)$ in powers of $\Phi$. At order $\Phi^4$
we can write the expansion as 
\begin{eqnarray}
W(\Phi) &=& a_{11} \Phi_z + a_{21} (\Phi^2_x + \Phi^2_y) + a_{22} \Phi^2_z
      + a_{31} \Phi^3_z + a_{32} \Phi_z (\Phi^2_x + \Phi^2_y) 
\nonumber \\ 
      && + a_{41} (\Phi^2_x + \Phi^2_y)^2 + 
        a_{42} (\Phi^2_x + \Phi^2_y) \Phi^2_z + 
        a_{43} \Phi^4_z + O(\Phi^5),
\label{expanW}
\end{eqnarray}
where $a_{ij}$ can be easily computed in terms of $H$ and
$D$. Finally, we identify the critical modes that are associated with
the wavevectors $Q$ that maximize the kinetic term. In our case the
relevant modes are associated with the wavevector $Q =
(4\pi/3,0,\pi)$, so that we can write
\begin{eqnarray}
\Phi_x(r) &=& c[ \phi_{11}(r) \cos (Q\cdot r) + 
                 \phi_{12}(r) \sin (Q\cdot r)], \nonumber \\
\Phi_y(r) &=& c[ \phi_{21}(r) \cos (Q\cdot r) + 
                 \phi_{22}(r) \sin (Q\cdot r)], \nonumber \\
\Phi_z(r) &=& c[ \psi_{1}(r) \cos (Q\cdot r) + 
                 \psi_{2}(r) \sin (Q\cdot r)],
\label{crit-modes}
\end{eqnarray}
where we have introduced new fields $\phi_{ai}$ and $\psi_i$, with
$a=1,2$ and $i=1,2$.  The constant $c$ is fixed so that, keeping only
slowly varying contributions, we have
\begin{equation}
    {1\over 2\beta} \sum_{ij} (J^{-1})_{ij} \Phi_i\cdot \Phi_j \approx 
        - \int d^3x\, \left[
         {1\over2} \sum_\mu [\sum_{ai} (\partial_\mu \phi_{ai})^2 + 
               \sum_i (\partial_\mu \psi_i)^2] +
        {1\over 2} a_0 (\phi^2 + \psi^2) \right]. 
\end{equation}
Substituting expressions (\ref{crit-modes}) in Eq.~(\ref{expanW}) and
keeping only slowly varying contributions, we obtain the Hamiltonian
(\ref{LGWH}) with $r_\phi = (c^2 a_{21} + a_0)$, $r_\psi = (c^2 a_{22}
+ a_0)$, $u_0 = 9 a_{42} c^4/2$, $v_0 = 3 a_{42} c^4$, $w_0 = 9 (2
a_{41} - a_{42}) c^4/2$, $y_0 = 9 (2 a_{43} - a_{42}) c^4/2$, and $z_0
= 3 (2 a_{41} - a_{42}) c^4$.  This derivation gives relations among
the different couplings appearing in Eq.~(\ref{LGWH}).  However, they
should not be taken seriously, since the effective Hamiltonian
(\ref{LGWH}) is only an approximation of the original one.  Note that
the odd powers of $\Phi$ present in Eq.~(\ref{expanW}) do not
contribute to the effective Hamiltonian. Indeed, here the basic ingredient
is the anisotropy, i.e. the breaking of the O(3) spin symmetry to
${\mathbb Z}_2\oplus$O(2). The additional breaking of the ${\mathbb
Z}_2$ symmetry caused by the magnetic field does not play any role,
apart from modifying the explicit expressions of the parameters of the
LGW Hamiltonian.

Several comments should be made on the derivation of the effective
Hamiltonian (\ref{LGWH}). First, we have assumed here that the
relevant modes can be inferred from the analysis of the hopping term.
While this is correct for unfrustrated systems, for frustrated ones as
is the case here the method is questionable and it is indeed possible
that some low-temperature properties are not correctly described by
this approach \cite{Coppersmith_85}. Second, note that the explicit
value of $Q$ does not play any role. This means that the effective
Hamiltonian (\ref{LGWH}) can be used to describe any critical behavior
with a three-component order parameter $S(q)$, with $q \not=K/2$,
where $K$ is a reciprocal-lattice vector. The effective Hamiltonian
(\ref{LGWH}) differs from that presented in
Refs.~\cite{PHC-88,KCP-90}, in which the quartic breaking terms
proportional to $w_0$, $y_0$, and $z_0$ are absent.\footnote{
Hamiltonian (\ref{LGWH}) is recovered by adding
in the Ansatz of Ref.~\cite{PHC-88} for the free energy,
see their Eq.~(3),
terms proportional to $|S_z|^4$, $S^2 |S_z|^2$,  and
$(S_z^2 S^*\cdot S^* + (S^*_z)^2 S\cdot S)$.
These terms are {\em a priori}
expected in the presence of easy-axis anisotropy
on the basis of symmetry considerations.  }
In any case, even if absent in the microscopic
model, these additional terms would be generated by RG
transformations.

\begin{figure}[t]
\vskip 1truecm
\centerline{\epsfig{file=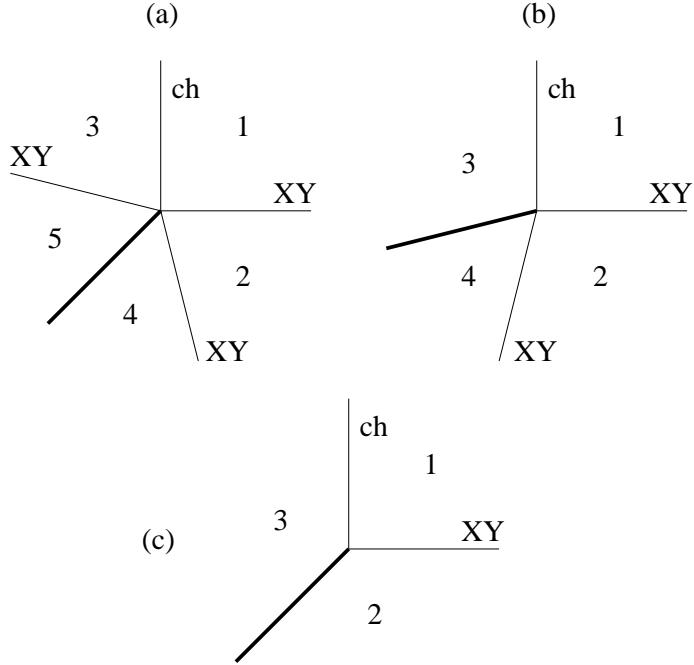,width=10truecm}}
\vskip -1.2truecm
\caption{The possible phase diagrams in the $(r_\phi,r_\psi)$ plane 
predicted by the mean-field approximation with a chiral ("ch") transition.
Thin lines represent second-order transitions, while thick lines 
are first-order transitions. Phase 1 is paramagnetic, in phase 2 
$\psi \not =0$ and $\phi = 0$, in phase 3 
$\psi = 0$, $\phi_1\not=0$, $\phi_2\not=0$ with $\phi_1 \cdot \phi_2 = 0$,
in phase 4 $\phi_1 \not = 0$, $\phi_2 = 0$, $\psi \not = 0$, 
in phase 5 all vectors are nonvanishing.}
\label{fig-MF}
\end{figure}

The phase diagram of the theory with Hamiltonian (\ref{LGWH}) can be
studied within the mean-field approximation. The general discussion is
presented in App.~\ref{AppA}. Here we only report the final results
for the specific case in which one of the transitions is a chiral
transition, i.e. it is associated with the symmetry-breaking pattern
O(2)$\otimes$O(2)$\to$O(2)$_{\rm diag}$, as it is of interest for
easy-axis materials.  This occurs when $v_0 + z_0 > 0$.  The reader
interested in systems with a collinear/paramagnetic transition is
referred to App.~\ref{AppA}. The possible phase diagrams are reported
in Fig.~\ref{fig-MF}.  There are three possibilities: (a) a
pentacritical point, (b) a tetracritical point, (c) a bicritical
point.  In all cases but one the transitions are second-order ones;
one transition line is of first order. The known easy-axis materials, like
ANiX$_3$, with A = Cs, Rb, and X = Cl, Br, or CsMnI$_3$, all show a
tetracritical point, i.e. a phase diagram of type (b). This fact
should be related to the smallness of the easy-axis anisotropy.
Indeed, for $D\to 0$ we also have $H\to 0$ at the multicritical point.
Thus, the breaking of the O(3) invariance is expected to be small at
the multicritical point, and we can generically assume that $w_0
\approx y_0 \approx z_0 \approx 0$.  In this specific case, the
mean-field analysis predicts a tetracritical phase diagram
irrespective of $u_0$ and $v_0$, in agreement with
experiments. Bicritical behavior is expected for $|D|$ large enough:
for $|D| > 3 J'$, where $J'$ is the intraplane coupling, the basal
spin components $s_x$ and $s_y$ should not magnetize at any
temperature \cite{CP-97} and thus phase 4 should not occur, forbidding
a tetracritical behavior.  For $H = 0$, these systems should behave as
Ising antiferromagnets.  Note that, in this case, beside the XY
transition predicted by the phase diagram (c), other transitions
(probably first-order ones) may occur as $T$ is lowered
\cite{BMNBGS-84,NBGSBM-84,PCH-89}.  It is interesting to note that the
mean-field analysis also predicts a pentacritical point. In this case
there is a new phase (phase 5 in Fig.~\ref{fig-MF}) in which the basal
spin components $s_x$ and $s_y$ show a distorted 120$^\circ$
structure, while the $s_z$ component is modulated as in phase 2.  The
possibility of a tetracritical phase diagram was already noted in
Ref.~\cite{PHC-88}. However, in their work, since they only considered
the case $w_0 = y_0 = z_0 = 0$, this was the only possible phase
diagram.  Indeed, the pentacritical and the bicritical points are
only obtained if $w_0$, $y_0$, and $z_0$ are not all vanishing.

\section{Analysis of some particular cases}
\label{sec3}

\subsection{Particular models and fixed points}
\label{sec3.1}

The three-dimensional properties of the RG flow are determined by its
FPs.  Some of them can be identified by considering particular cases
in which some of the quartic parameters vanish.  For example, we can
easily recognize:
\begin{itemize}
\item[(a)] 
the O($K$)-symmetric model is recovered by setting
$r_\phi=u_0=v_0=w_0=z_0=0$ ($K=2$), $r_\psi=u_0=v_0=y_0=z_0=0$
($K=2N-2$), $r_\phi=r_\psi$ and $v_0=w_0=y_0=z_0=0$ ($K=2N$).  Results
for these theories are reviewed, e.g., in Ref.~\cite{PV-rev}.
\item[(b)] 
the O(2)$\otimes$O($K$) model, cf.~Eq.~(\ref{LGWH-chirale})
with $\Phi_{ai}$ being a $K\times2$ matrix, for $r_\phi=r_\psi$ and
$w_0=y_0=z_0=0$ ($K=N$), and for $r_\psi=0$ and $u_0=v_0=y_0=0$
($K=N-1$).  The properties of these models are reviewed in
Refs.~\cite{Kawamura-98,CPPV-04,PV-rev,DMT-03,DPV-03}.  In three
dimensions perturbative calculations within the 
MZM scheme \cite{PRV-01,CPS-02} and within the $3d$-$\overline{\rm
MS}$ scheme \cite{CPPV-04} indicate the presence of a stable FP with
attraction domain in the region $g_{2,0}>0$ for all values of $K$,
except possibly $K=6$.  For $K=2$, these conclusions have been
recently confirmed by a Monte Carlo calculation \cite{CPPV-04}.  On
the other hand, near four dimensions, a stable FP is found only for
large values of $K$, i.e., $K> K_c= 21.80 - 23.43 \epsilon + 7.09
\epsilon^2 + O(\epsilon^3)$ \cite{Kawamura-88,ASV-95,PRV-01b,CP-03}.
A stable FP with attraction domain in the region $g_{2,0}<0$ exists
for $K=2$ (it belongs to the XY universality class)
\cite{Kawamura-88}, for $K=3$ (Ref.~\cite{DPV-03}), and, as we shall
discuss below, for $K=4$.  Note that nonperturbative approximate RG
calculations have so far found no evidence of stable FPs for $K=2$ and
3 \cite{DMT-03,KW-01}. In the following we will call the FP with $g_2>
0$ {\em chiral} FP, while the FP with $g_2 < 0$ will be named {\em
collinear} FP.

\item[(c)] 
the O(2)$\oplus$O($K$) model with $K=2N-2$ for $v_0=z_0=0$.
This theory describes the multicritical behavior of a model with two
order parameters that is symmetric under the group O(2)$\oplus$O($K$)
\cite{KNF-76}.  In the case we are interested in, i.e. for $N\geq 3$
and therefore $K\geq 4$, the stable FP is the decoupled FP,
corresponding to $u_0=0$, which describes a critical behavior in which
the two order parameters $\phi$ and $\psi$ are effectively uncoupled
at the multicritical point \cite{CPV-03,Aharony-02}.

\item[(d)]
Decoupled O(2)$\otimes$O($N-1$) and O(2) models for
$u_0=v_0=0$. The corresponding stable FP describes two effectively
decoupled critical behaviors in the O(2)$\otimes$O($N-1$) and O(2)
universality classes.

\item[(e)] 
For $v_0 = 0$ we obtain a multicritical theory with the
larger symmetry group [O(2)$\otimes$O($N-1$)]$\oplus$O(2).
\end{itemize}
The FPs of the above-mentioned particular models are also FPs of the enlarged
model (\ref{LGWH}). Their stability in the full theory can be checked 
by computing the RG dimensions of the 
additional terms present in the complete Hamiltonian
(\ref{LGWH}). 

The analysis of model (\ref{LGWH}) is simplified by the following
symmetry transformation. If we transform the fields as
\begin{equation}
   \psi_i \to \sum_j \epsilon_{ij} \psi_j, \qquad 
   \phi_{ai} \to \phi_{ai},
\end{equation}
and the couplings according to 
\begin{equation}
u_0\rightarrow u_0-v_0, \qquad v_0\rightarrow -v_0,\qquad
w_0\rightarrow v_0+w_0, \qquad y_0\rightarrow v_0+y_0,\qquad 
z_0\rightarrow 2 v_0+z_0,
\label{sym1}
\end{equation}
we reobtain the same Hamiltonian. This transformation leaves invariant
the model with $v_0 = 0$ and maps any model with $v_0 < 0$ to a model
with $v_0 > 0$. The transformation (\ref{sym1}) implies several
symmetry properties for the RG functions. In particular, under the
transformation of the renormalized quartic couplings
\begin{equation}
u\rightarrow u-v, \qquad v\rightarrow -v,\qquad
w\rightarrow v+w, \qquad y\rightarrow v+y,\qquad 
z\rightarrow 2 v+z,
\label{sym}
\end{equation}
the RG functions associated with the exponents are unchanged, while
the $\beta$-functions transform covariantly.  As a consequence, each
FP with nonvanishing quartic coupling $v < 0$ is mapped into another
FP with $v> 0$ and the same stability properties.  The symmetry
(\ref{sym1}) implies that there exists another O(2)$\otimes$O($N$)
model beside that reported at point (b) above:
\begin{itemize}
\item[(b$^\prime$)] The O(2)$\otimes$O($N$) model is also obtained for 
$r_\phi=r_\psi$, 
$w_0 = - v_0$, $y_0 = - v_0$, $z_0 = - 2 v_0$;
in this case $g_{1,0} = u_0 - v_0$, $g_{2,0} = - v_0$.
\end{itemize}
Because of symmetry (\ref{sym}) it is enough to study the RG flow 
for $v \ge 0$.

The results for models (a), (b), (b$^\prime$), (c), and (d) allow us
to identify four possible FPs that are candidates for being stable FPs
of the full theory.  In the O(2)$\otimes$O($N$) model
(\ref{LGWH-chirale}) there may be, depending on the value of $N$, two
FPs. The chiral FP is located at $g_1 = g^*_{1,\rm ch}(N)$ and $g_2 =
g^*_{2,\rm ch}(N)$ with $g^*_{2,\rm ch}(N)> 0$ and exists for any
value of $N$ except possibly $N=6$ \cite{CPPV-04}.  The collinear FP
is located at $g_1 = g^*_{1,\rm cl}(N)$ and $g_2 = g^*_{2,\rm cl}(N)$
with $g^*_{2,\rm cl}(N) < 0$. Such a FP exists for $N=2$
\cite{Kawamura-88}---in this case it is equivalent to the standard XY
FP---and for $N=3$ \cite{DPV-03}.  We have investigated if the
collinear FP exists also for larger values of $N$, by extending the
analysis of Ref.~\cite{DPV-03}.  In the $3d$-$\overline{\rm MS}$
scheme at five loops, we find a stable FP only for $N=4$: $g^*_{1,\rm
cl}(4) = 0.10(7)$ and $g^*_{2,\rm cl}(4) = -1.83(10)$.\footnote{
In the $3d$-$\overline{\rm MS}$ scheme the couplings are normalized
so that $g_i = g_{i,0} \mu^{-\epsilon}/A_d$ with
$A_d = 2^{d-1} \pi^{d/2} \Gamma(d/2)$.
In the MZM scheme the renormalized couplings $g_1$ and $g_2$ are normalized
so that $g_i = g_{i,0}/m$ at tree level,
where $m$ is the renormalized mass.
For reference, we report the collinear FP for $N=3$ in these
normalizations (Ref.~\cite{DPV-03}):
$g_{1,\rm cl}^*=-7.0(5)$ and $g_{2,\rm cl}^*=-50(2)$ in the
MZM scheme,
$g_{1,\rm cl}^*=0.04(8)$ and $g_{2,\rm cl}^*=-1.71(9)$
in the $3d$-$\overline{\rm MS}$ scheme.
The same normalization is used for the $\overline{\rm MS}$
couplings in Sec.~\ref{sec4.1}. }
For $N=5$, we find a FP only in 1/3 of the
cases that are considered and such a percentage decreases as $N$
increases.  In the MZM scheme at six loops a FP is found for $N=4$, 5,
and 6, and disappears for $N\ge 8$. For $N=4$ it is located
at $g^*_{1,\rm cl}(4) = 6.1(4)$ and
$g^*_{2,\rm cl}(4) = -50(2)$; for $N=5$ and $N=6$ at
$g_{1,\rm cl}^*=-7.2(5)$ and $g_{2,\rm cl}^*=-53(3)$,
$g_{1,\rm cl}^*=-8.5(7)$ and $g_{2,\rm cl}^*=-56(4)$, respectively.
The perturbative
analysis provides therefore strong evidence for the existence of a
collinear FP for $2\le N \le 4$. For $N\ge 8$ this FP is absent (in
agreement with the large-$N$ analysis), while in the intermediate
cases $5\le N \le 7$ it is not clear whether the collinear FP really
exists since the two perturbative schemes give opposite results.  We
have also verified the stability of the collinear FPs within the
O(2)$\otimes$O($N$) theory (\ref{LGWH-chirale}): whenever they exist,
they are stable. However, for $N=4$ we have been unable to estimate
the stability eigenvalues. Indeed, the $3d$-${\overline{\rm MS}}$
scheme predicts complex stability eigenvalues, while in the MZM scheme
we find complex eigenvalues only in 50\% of the cases.

The collinear FP has $g_2<0$. Therefore, if we are only interested in
the full model for $v \ge 0$, we must consider the FP appearing in
model (b$^\prime$). Thus, the above-reported results for the
O(2)$\otimes$O($N$) predict two possible FPs:
\begin{itemize}
\item[(a)] $u = g^*_{1,\rm ch}(N)$, $v = g^*_{2,\rm ch}(N)$, $w=y=z=0$;
\item[(b)] $u = g^*_{1,\rm cl}(N) - g^*_{2,\rm cl}(N)$, 
   $w = y = z/2 = -v = g^*_{2,\rm cl}(N)$.
\end{itemize}
Analogously, analysis of the decoupled theory gives two possible 
FPs:
\begin{itemize}
\item[(c)] $u=v = 0$, $w = g^*_{1,\rm ch}(N-1)$, 
           $z = g^*_{2,\rm ch}(N-1)$, $y = g^*_{\rm XY}$;
\item[(d)] $u=v = 0$, $w = g^*_{1,\rm cl}(N-1)$, 
           $z = g^*_{2,\rm cl}(N-1)$, $y = g^*_{\rm XY}$.
\end{itemize}
Here $g^*_{\rm XY}$ is the four-point renormalized coupling of the XY theory
(numerical estimates can be found in Ref.~\cite{CHPRV-01}). 
Note that FPs (c) and (d) are also FPs of the theory with $v=0$.

\subsection{Stability of the O(2)$\otimes$O($N$) fixed points}
\label{sec3.2}

In this section we study the stability properties of the two FPs that
appear in the O(2)$\otimes$O($N$) model and, therefore, we check the
possibility of an enlargement of the symmetry at the multicritical
point from O(2)$\otimes$[${\mathbb Z}_2\oplus$O($N-1$)] to
O(2)$\otimes$O($N$). For this purpose we need to classify the
perturbations that break the O($N$) symmetry to O($N-1$) and preserve
the O(2) symmetry, according to their transformation properties under
the O($N$) group and the number of fields.  This is done in detail in
App.~\ref{AppB}.  The multicritical Hamiltonian (\ref{LGWH}) can be
rewritten as
\begin{equation}
{\cal H} = {\cal H}_{\rm sym} + r_2 V^{(2,2)} + f_1 V^{(4,4)} + 
      f_2 V^{(4,2,1)} + f_3 V^{(4,2,2)},
\end{equation}
where ${\cal H}_{\rm sym}$ is the O(2)$\otimes$O($N$)-symmetric
Hamiltonian (\ref{LGWH-chirale}) obtained by setting $w_0=y_0=z_0=0$
and $r_\psi=r_\phi$, and $V^{(2,2)}$, $V^{(4,4)}$, $V^{(4,2,1)}$, and
$V^{(4,2,2)}$ are respectively a spin-2 quadratic term, a spin-4
quartic term, and two spin-2 quartic terms. Their explicit expressions
can be found in App.~\ref{AppB}.  The spin-2 quadratic pertubation
$V^{(2,2)}=\phi^2/N - (N-1)\psi^2/N$ is always relevant.  Its RG
dimension at the O(2)$\otimes$O($N$) FP gives the crossover exponent,
i.e. $\phi=\nu y_{2,2}$, where $\nu$ is the correlation-length
exponent.

Let us first discuss the chiral FP (a) that has $g^*_{2,\rm ch}(N) >
0$.  The exponent $y_{2,2}$ coincides with the exponent $y_3$ defined
in App.~\ref{AppC}, since $V^{(2,2)} = \sum_{k=1}^{N-1} O^{(3)}_{kk}$.
For $N=3$, we obtain $y_{2,2} = 1.49(3)$ in the MZM scheme and
$y_{2,2} = 1.54(8)$ in the $3d$-$\overline{\rm MS}$ scheme.

In order to estimate the RG dimensions $y_{4,2,1}$, $y_{4,2,2}$, and
$y_{4,4}$ of the quartic perturbations, we computed the corresponding
five-loop series in the $\overline{\rm MS}$ scheme and we analyzed
them within the $3d$-$\overline{\rm MS}$ scheme.  Note that, since the
spin-2 quartic operators mix, $y_{4,2,1}$ and $y_{4,2,2}$ are the
eigenvalues of the corresponding RG-dimension matrix.  We also
estimated the spin-4 RG dimension $y_{4,4}$ by computing the
corresponding five-loop series in the MZM scheme.  In this scheme we
cannot estimate the RG dimensions of the spin-2 quartic perturbations
because they also mix with the lower-dimension spin-2 operator
$V_{2,2}$. Such a mixing does not occur in the $3d$-$\overline{\rm
MS}$ scheme, since in this case the theory is massless and, therefore,
operators of different naive dimensions do not mix under
renormalization.

\begin{table}
\squeezetable
\caption{ 
Coefficients $a_{ij}$ of the five-loop $\overline{\rm MS}$ and MZM
expansions of $y_{4,4}$, cf. Eqs.~(\ref{y44ms}) and (\ref{y44mzm}),
for $N=3$ and 4.
}
\begin{tabular}{crrrr}
\multicolumn{1}{c}{$$}& 
\multicolumn{2}{c}{$\overline{\rm MS}$}&
\multicolumn{2}{c}{MZM}\\
\multicolumn{1}{c}{$i,j$}&
\multicolumn{1}{c}{$N=3$}& 
\multicolumn{1}{c}{$N=4$}& 
\multicolumn{1}{c}{$N=3$}& 
\multicolumn{1}{c}{$N=4$}\\ 
\hline
1,0 & $-$2     & $-$2  & $-$6/7 &  $-$3/4  \\
0,1 & $-$1/3   & $-$1/3   & $-$2/9  & $-$2/9   \\
2,0 & 28/9     & 61/18  & 0.383976  &  0.320602 \\
1,1 & 1/9      & $-$1/6  & 0.018812  & $-$0.030864    \\
0,2 & $-$1/36  & 1/8 & $-$0.005487    &  0.041152  \\
3,0 & $-$11.157290 & $-$12.122093  &$-$0.246143 & $-$0.176781  \\
2,1 & $-$0.245370  & 0.984242 & $-$0.016559 &  0.015113 \\
1,2 & $-$1.174531  & $-$2.872196 & $-$0.072221 & $-$0.142869  \\
0,3 & 0.824994  & 1.509014 &0.080442& 0.140829  \\
4,0 & 62.535697 & 72.040868 &0.225243  & 0.152346  \\
3,1 & $-$6.135544  & $-$19.807535 & $-$0.038487 &$-$0.083599   \\
2,2 & 14.290714  & 31.675613 &0.157437  &0.252196   \\
1,3 & $-$8.768847  & $-$16.889435 & $-$0.138290 & $-$0.224249 \\
0,4 & 0.999971  & 1.877179 & 0.008548  &  0.019106 \\
5,0 & $-$422.234940 & $-$508.920947  & $-$0.231515 & $-$0.138376  \\
4,1 & 90.411170 &  230.058460 & 0.069042 & 0.102807 \\
3,2 & $-$166.086612  & $-$359.525801 & $-$0.268436 & $-$0.357188  \\
2,3 & 108.042417  & 220.680543 & 0.271663 &  0.388611 \\
1,4 & $-$23.261600  & $-$49.094226 & $-$0.061312 & $-$0.100610  \\
0,5 & 2.329711  & 4.789714  & 0.015950 & 0.019262  \\
\end{tabular}
\label{coeffy44}
\end{table}

Here we only report the series for $y_{4,4}$, which will be the most
relevant for the analysis of the stability of the O(2)$\otimes$O($N$)
FP, for $N=3$ and 4, and in the $\overline{\rm MS}$ and MZM schemes.  In the
$\overline{\rm MS}$ scheme we have
\begin{eqnarray}
y_{4,4}= \epsilon + \sum_{ij} a_{ij} g_1^i g_2^j,
\label{y44ms}
\end{eqnarray}
with $\epsilon = 1$ in three dimensions.
The coefficients $a_{ij}$ for $N=3,4$ are reported in Table \ref{coeffy44}
to five loops, i.e. for $i+j\le 5$.
The renormalized $\overline{\rm MS}$ couplings $g_{1}$ and $g_2$, corresponding to the 
quartic parameters $g_{1,0}$ and $g_{2,0}$ of the Hamiltonian (\ref{LGWH-chirale}),
are normalized as in Ref. \cite{CPPV-04}, see footnote 2.
In the MZM scheme we have the analogous expansion\footnote{
The perturbative MZM series were obtained by using a symbolic
manipulation program that generated the diagrams, the symmetry, and
group factors. Numerical estimates of the Feynman integrals
were taken from Ref.~\cite{NMB-77}. }
\begin{eqnarray}
y_{4,4}= 1 + \sum_{ij} a_{ij} \bar{g}_1^i \bar{g}_2^j
\label{y44mzm}
\end{eqnarray}
The coefficients $a_{ij}$ to five loops are reported in Table
\ref{coeffy44} for $N=3,4$.  The renormalized MZM couplings
$\bar{g}_{1,2}$ are normalized as in Ref. \cite{PRV-01},
i.e. $\bar{g}_i\approx c_i g_{i,0}/m$ at tree order, where
$c_1=(8+N)/(48\pi)$ and $c_2=3/(16\pi)$.  The perturbative series 
that are not reported here are available on request.

We analyzed the series using the conformal-mapping method and the
Pad\'e-Borel method, following closely Refs.~\cite{CPV-00,CPPV-04}, to
which we refer for details.  The error on the conformal-method results
takes into account the spread of the results as the parameters
$\alpha$ and $b$ are varied (cf.~Ref.~\cite{CPV-00} for definitions)
and the error due to the uncertainty of the FP location (we use the
estimates reported in Refs.~\protect\cite{PRV-01,CPS-02,CPPV-04,DPV-03}). In
the Pad\'e-Borel analysis we used the Pad\'e [4/1] and several values
of the parameter $b$. Again, the error takes into account the
dependence on $b$ and the uncertainty of the FP location.  In the
Pad\'e-Borel analyses of $y_{4,4}$ the error also takes into account the 
difference between the [4/1] and the [3/1] estimates.  This was not done for 
the spin-2 RG dimensions since in that case the results obtained by using
the Pad\'e [3/1] did not look reliable.  They varied significantly
with $b$ and, for $N\le 5$, favored complex estimates of $y_{4,2,1}$
and $y_{4,2,2}$.

\begin{table}
\squeezetable
\caption{ Estimates of the RG dimensions $y_{4,2,1}$, $y_{4,2,2}$, and
$y_{4,4}$ of the operators $V^{(4,2,1)}$, $V^{(4,2,2)}$, and
$V^{(4,4)}$ at the chiral (ch) FP and at the collinear (cl) FP.
Results have been obtained in the $3d$-$\overline{\rm MS}$ scheme
($\overline{\rm MS}$) and in the three-dimensional massive
zero-momentum scheme (MZM).  For the resummation of the five-loop
perturbative series the conformal-mapping (CM) and the Pad\'e-Borel
(PB) methods have been used.  For $y_{4,4}$ two estimates are reported
in each case, resspectively from the analysis of the series
of $y_{4,4}$ and $1/y_{4,4}$.
}
\begin{tabular}{clllllll}
\multicolumn{1}{c}{FP,$N$}& 
\multicolumn{2}{c}{$y_{4,2,1}$}&
\multicolumn{2}{c}{$y_{4,2,2}$}&
\multicolumn{3}{c}{$y_{4,4}$} \\
\multicolumn{1}{c}{}& 
\multicolumn{1}{c}{$\overline{\rm MS}$, CM}&
\multicolumn{1}{c}{$\overline{\rm MS}$, PB}&
\multicolumn{1}{c}{$\overline{\rm MS}$, CM}&
\multicolumn{1}{c}{$\overline{\rm MS}$, PB}&
\multicolumn{1}{c}{$\overline{\rm MS}$, CM}&
\multicolumn{1}{c}{$\overline{\rm MS}$, PB}&
\multicolumn{1}{c}{MZM, CM} \\
\hline
ch,3 &  $-$0.7(9) & $-$1.8(7) & 0.0(7) & $-$0.9(6) & 
$-$0.4(5), 0.12(5) & $-$0.4(4), 0.07(3)  &  $-$0.3(2), 0.05(2) \\ 

ch,4  & $-$0.7(5) & $-$1.0(3) & 0.3(3) & $-$0.2(3) & 
$-$0.1(2), 0.18(4)  & $-$0.2(2), 0.12(4) &  0.15(12), 0.16(4) \\ 

ch,5  & $-$0.8(7) & $-$0.6(3) & 0.2(2) & $\phantom{-}$0.0(2) & 
0.0(2), 0.22(4)  & 0.02(11), 0.18(5)  & 0.3(2), 0.24(5) \\ 

ch,6  & $-$0.6(6) & $-$0.5(2) & 0.25(13) & $\phantom{-}$0.11(15) & 
 0.1(2), 0.26(4) &  0.13(10), 0.23(5) & \\ 

ch,8  & $-$0.5(5) & $-$0.4(2) & 0.22(9) & $\phantom{-}$0.17(6) & 
0.2(2), 0.34(5) & 0.29(8), 0.32(4) & 0.26(2), 0.29(2) \\ 

ch,16  & $-$0.3(2) & $-$0.17(6) & 0.09(2) & $\phantom{-}$0.08(3) & 
0.56(4), 0.570(9) & 0.60(2), 0.58(3) & 0.56(2), 0.54(2) \\ 

ch,$\infty$ & \multicolumn{2}{c}{0} & 
              \multicolumn{2}{c}{0} & 
              \multicolumn{3}{c}{1} \\

cl,3 & & & & & 
0.3(4), 0.7(2) &  $-$0.7(1.1), 0.5(2) &  0.5(1.0), 1.2(7) \\ 
cl,4 & & & & & 
$-$0.2(8), 0.5(2) & $-$1(1), 0.35(15) &  0.5(1.1), 0.9(3) \\ 
\end{tabular}
\label{tab2}
\end{table}

The results of the analyses are reported in Table~\ref{tab2}.  
We first computed the spin-2 RG dimensions $y_{4,2,1}$ and $y_{4,2,2}$.
If $Y_{ij}$, $j=1,2$, is the anomalous-dimension matrix,
$y_{4,2,1}$ and $y_{4,2,2}$ are the eigenvalues of $Y$.
In order to determine them, we resummed the elements $Y_{ij}$
and computed the two eigenvalues. The results we report are obtained by
averaging the eigenvalues over many different choices of approximants.  
The results have a quite large error,
so that it is impossible to draw definite
conclusions on the relevance of these operators. For $N\ge 5$ the
results favor $y_{4,2,2} > 0$, so that one operator would be relevant,
while for $N=3$ it seems likely that the spin-2 perturbations are
irrelevant.
For the spin-4 RG dimension we report two estimates for each
case. They are obtained from the analysis of the series of $y_{4,4}$
and $1/y_{4,4}$, respectively.
Their difference should allow us to estimate systematic
errors in the series resummations that are not taken into
account by the spread of the approximants, which we usually take as 
an indication of the error.  We find
$y_{4,4} > 0$ for all $N\ge 4$, with increasing confidence as $N$
increases (for $N=4$ we mainly rely on the MZM analysis that predicts
$y_{4,4} > 0$ both for the direct and for the inverse series).  For
$N=3$, the results of the analysis of $y_{44}$ and $1/y_{44}$ 
differ substantially (the error of the estimate from $1/y_{44}$
may be underestimated) and even have opposite signs. This does not allow
us to establish whether $V^{(4,4)}$ is a relevant or an irrelevant
perturbation of the chiral FP. 

Let us briefly discuss the physical picture in the two cases.
If the O(2)$\otimes$O(3) FP is stable, it controls the
critical behavior of statistical systems in its attraction domain.
Setting $t\equiv (T-T_{mc})/T_{mc}$, where $T_{mc}$ is the critical temperature
at the multicritical point, the singular part of the free energy can
be written as
\begin{equation}
{\cal F}= |t|^{3\nu}f(A|t|^{-\phi}),
\label{freeen}
\end{equation}
where $A$ is the scaling field associated with the anisotropy---in STA's it 
will be a combination of $T$, $D$, and $H$---$\nu\approx 0.6$ is the 
correlation-length exponent of the
O(2)$\otimes$O(3) theory \cite{PRV-01,CPPV-04}, $\phi=y_{2,2}\nu\approx
0.9$ is the crossover exponent, and $f(x)$ is a scaling function.  

If the O(2)$\otimes$O(3) is unstable with respect to the
spin-4 quartic perturbation, one should also consider the
crossover exponent $\phi_{4,4}$ associated with the spin-4 quartic
instability. This is expected to be quite small.  Indeed, our
results indicate $y_{4,4}\lesssim 0.2$ for $N=3$, so that
$\phi_{4,4}\lesssim 0.1$.  Therefore, if the O(2)$\otimes$O(3) FP is
unstable and the effective quartic Hamiltonian parameters breaking the
O(2)$\otimes$O(3) symmetry are small, the crossover from the
preasymptotic O(2)$\otimes$O(3) critical behavior to the eventual
asymptotic behavior is expected to be very slow and one may observe an
effective O(2)$\otimes$O(3) critical behavior for a wide range of
reduced-temperature values.

Now, let us consider the collinear FP (b) that has $g_{2,\rm cl}^*(N)
< 0$ for $N=3$ and $N=4$. For the spin-2 operators the analysis
indicates that they are irrelevant at the FP, but is not precise
enough to allow a quantitative determination of $y_{4,2,1}$ and
$y_{4,2,2}$.  For $N=3$, by using the conformal-mapping method we find
that approximately 50\% of the approximants give complex spin-2 RG
dimensions (the real part is always negative)
and 50\% provide real negative estimates.
Pad\'e-Borel approximants 
always give real negative estimates. 
The same pattern is observed for $N=4$. The results for the 
spin-4 operator are reported in Table~\ref{tab2}. 
They suggest that the collinear FP is unstable for both $N=3$ and $N=4$.

In conclusion, the perturbative analysis shows that the collinear FP
is always unstable, while the chiral FP is unstable for $N\ge 4$.  For
$N=3$ the results are not conclusive and the O(2)$\otimes$O(3) chiral
FP may be either stable or unstable.

\subsection{Stability of the decoupled [O(2)$\otimes$O($N-1$)]$\oplus$O(2) 
fixed points}
\label{sec3.3}

Two other interesting FPs can be investigated by essentially
nonperturbative arguments, FPs (c) and (d) discussed in
Sec.~\ref{sec3.1}.  In order to check the stability of these FPs, we
must determine the RG dimensions at the decoupled FPs of the
perturbations
\begin{equation}
P_E=\int d^3x\, \phi^2 \psi^2, \qquad 
P_T=\int d^3x\, \sum_{ij} O^{(4)}_{ij} T_{ij},
\end{equation}
where $O^{(4)}_{ij}$ is defined in App.~\ref{AppC}, cf.~Eq.~(\ref{o4})
and $T_{ij} = \psi_i \psi_j - {1\over2} \delta_{ij} \psi^2$.
Simple RG arguments show that the RG dimension of $P_E$,
which is an energy-energy term, is given by
\begin{equation}
y_E = 
{1\over \nu_\psi} + {1\over \nu_\phi} - 3 = 
{\alpha_\psi \over 2 \nu_\psi} + 
{\alpha_\phi \over 2 \nu_\phi} , 
\label{ye}
\end{equation}
where $\alpha_\psi$ and $\nu_\psi$ are the critical exponents of the
3-dimensional XY universality class ($\alpha_\psi = -0.0146(8)$ and
$\nu_\psi=0.67155(26)$, see Ref.~\cite{CHPRV-01}), while $\alpha_\phi$
and $\nu_\phi$ are those of the 3-dimensional theory associated with
the O(2)$\otimes$O($N-1$) FP.  Analogously, for the RG dimension of
$P_T$ we obtain the relation
\begin{equation}
y_T = y_{T,\psi} + y_{4,\phi} - 3,
\label{yt}
\end{equation}
where $y_{T,\psi}=y_{T,{\rm XY}}$, $y_{T,{\rm XY}}$
is RG dimension of the operator
$T_{ij}$ at the XY FP
($y_{T,{\rm XY}}\approx 1.77$ \cite{CPV-02,CPV-04nova}), and $y_{4,\phi}$
is the RG dimension of the quadratic operator $O^{(4)}_{ij}$ introduced in
App.~\ref{AppC}, computed at the O(2)$\otimes$O($N-1$) FP. 

Let us consider first the chiral FP (c) that has $z > 0$.  Estimates
of the exponents $\alpha_\phi$ and $\nu_\phi$ for several values of
$N$ are reported in Refs.~\cite{PRV-01,CPS-02,CPPV-04}, while
estimates of $y_{4,\phi}$ are reported in App.~\ref{AppC}.  It is then
easy to check that $y_E>0$ for $N\lesssim 6$ and $y_E<0$ for $N\gtrsim
6$.  Therefore, the perturbation $P_E$ is relevant for $N\lesssim 6$.
Analogously, we obtain $y_T\approx 0$ for $N=3$ and $y_T<0$ for
$N>3$. Thus, except possibly for $N=3$, $P_T$ is always irrelevant.
These results indicate that the decoupled FP (c) is unstable for
$N\lesssim 6$, and therefore for the interesting case $N=3$.

Let us now consider the collinear FP (d). For $N=3$ the collinear FP
in the O(2)$\otimes$O(2) theory belongs to the XY universality class
\cite{Kawamura-88}.  It is easy to show that $\alpha_\phi =
\alpha_{\rm XY}$, which gives $y_E \approx -0.04$.  Thus $P_E$ is
irrelevant.  As for $P_T$, using the results of App.~\ref{AppC}, we
have $y_{4,\phi} = 2 y_{h,{\rm XY}} - 3$, where $y_{h,{\rm XY}} = (5 -
\eta_{\rm XY})/2$ is the RG dimension of the field in the XY
model. Therefore, in three dimensions $y_T = y_{T,{\rm XY}} -
\eta_{\rm XY} - 1 \approx 0.73$, which means that $P_T$ is relevant,
and the collinear decoupled FP (d) is unstable.  For $N=4$ and $N=5$,
using the results of App.~\ref{AppC}, we have $y_T > 0$ in both
cases. In order to compute $y_E$ we determined $\nu_\phi$ in the
O(2)$\otimes$O($N-1$) theory at the collinear FP. 
In the O(2)$\otimes$O(3) case we have
\cite{DPV-03} $\nu_{\phi} = 0.63(8)$ ($3d$-$\overline{\rm MS}$,
conformal mapping) and $\nu_\phi = 0.59(4)$ (MZM, conformal mapping);
in the O(2)$\otimes$O(4) case we have $\nu_{\phi} = 0.76(9)$
($3d$-$\overline{\rm MS}$, conformal mapping) and $\nu_\phi = 0.64(7)$
(MZM, conformal mapping).  These estimates imply $y_E > 0$ for $N=4$
and $-0.3 \lesssim y_E \lesssim 0.2$ for $N=5$.

In conclusion, the collinear decoupled FP is always unstable, while
the chiral decoupled FP is unstable for $N\lesssim 6$, stable in the
opposite case.

As we have mentioned in Sec.~\ref{sec3.1} the decoupled FPs are also
FPs for the multicritical theory with $v_0 = 0$. The analysis of their
stability in this particular case follows from the results reported
above.  Indeed, for the theory with $v_0=0$ it is enough to consider
the perturbation $P_E$. Thus, the decoupled chiral FP (c) is unstable for
$N\lesssim 6$ and stable in the opposite case. The decoupled collinear
FP (d) is stable for $N=3$ and possibly for $N=5$.

\section{The renormalization-group flow in the full theory} \label{sec4}

\subsection{Renormalization-group flow near four dimensions}
\label{sec4.1}

In Sec.~\ref{sec3} we considered some particular FPs, checking whether
they were stable in the full theory. In order to investigate the
presence of other FPs, we now perform a one-loop $\epsilon$-expansion
analysis.  The one-loop $\beta$ functions in the $\overline{\rm MS}$
scheme are:
\begin{eqnarray}
\beta_u&=&-\epsilon\,u+\frac{N+4}{3}u^2 -{N-1\over3}v \left(u-{v\over2}\right)
 +w\left(\frac{N}{3}u  - \frac{N-1}{6} v \right)
 \nonumber\\&&\qquad
 +y\left(\frac{2}{3}u -{v\over6}\right)- z{N-2\over6}(u-v),\\
\beta_v&=&  -\epsilon\,v +\frac{N-6}{6}v^2 + 2\,u\,v  + \frac{w\,v}{3} + 
   \frac{y\,v}{3} +  \frac{z\,(N-2)}{6}v ,\\
\beta_{w}&=&  -\epsilon\,w   + \frac{N+3}{3}w^2 
 -{N-2\over3}z \left(w-{z\over2}\right)
 +u\left( {N+6\over3} w  -\frac{2}{3} y + \frac{2-N}{6} z \right)
 \nonumber\\&&\qquad
 +v\left(-\frac{N-3}{6}w +\frac{1}{6} y+ \frac{N-2}{6}z  \right),\\
\beta_{y}&=& -\epsilon\,y   + \frac{5}{3} y^2
 +u\left(\frac{8}{3} y+ \frac{N-2}{6}z - \frac{N}{3}w\right) 
 +v\left( \frac{N-1}{6}w + \frac{y}{6} - \frac{N-2}{6}z \right),\\
\beta_{z}&=&  -\epsilon\,z   + 2\,w\,z+\frac{N-7}{6} z^2+ 
   2\,z\,u 
 +v\left(\frac{5}{3} w- \frac{y}{3} + \frac{N-12}{6} z \right),
\end{eqnarray}
where $u$, $v$, $w$, $y$, and $z$ are the 
$\overline{\rm MS}$ renormalized couplings 
(see footnote 2 for a precise definition).

For $N<N_c\approx21.8+O(\epsilon)$ one finds only six FPs (they belong
to the models considered in Sec.~\ref{sec3.1}) and none of them is
stable.  At $N = N_c$ four new FPs appear: they belong to the
O(2)$\otimes$O($N$) models considered in Sec.~\ref{sec3.1} [cf. cases
(b) and (b$^\prime$)].  It is easy to check that none of them is
stable, as it happens in three dimensions, cf. Sec.~\ref{sec3.2}.
Therefore, the enlargement of the symmetry to O(2)$\otimes$O($N$) is
never realized near four dimensions, for any value of $N$.  Stable FPs
are found only for $N>N_1\approx23.97$. For $N_1< N < N_2\approx24.15$
the stable FP has $v=0$ and does not belong to any of the models
considered in Sec.~\ref{sec3.1}.  For $N>N_2$ the only stable FP is
the decoupled chiral FP.  Note that, since the zeroes of the
$\beta$-functions are not degenerate for all $N\geq2$, no new FP can
emerge within the $\epsilon$ expansion at higher order.  The
conclusions on the stable FPs apply unchanged to the multicritical
theory with $v_0 = 0$ since the stable FPs always have $v = 0$.

This analysis can be extended to higher orders of the $\epsilon$
expansion, using the five-loop $\overline{\rm MS}$ series. On the
basis of our experience with the O(2)$\otimes$O($N$) model we expect
this analysis to be reliable only for large $N$. Indeed, in the
O(2)$\otimes$O($N$) theory the $\epsilon$ expansion is not able to
provide the correct three-dimensional scenario for the physically
interesting cases $N=2$ and 3, see, e.g., Ref.~\cite{CPPV-04}. Therefore,
we only studied the stability of the decoupled chiral FP, that,
according to the analysis presented in Sec.~\ref{sec3.3}, should be
stable for $N\gtrsim 6$.  For this purpose we computed the
stability-boundary function $N_2(\epsilon) \equiv n_2(\epsilon) + 1 =
24.15 + O(\epsilon)$.  Repeating the analysis of Sec.~\ref{sec3.3}
close to four dimensions, it is easy to show that $n_2(\epsilon)$ is
determined from
\begin{equation} 
    {1\over \nu_\phi(n_2,\epsilon)} + {1\over \nu_\psi(\epsilon)} = 
      4 - \epsilon,
\end{equation}
where $\nu_\phi(n_2,\epsilon)$ and $\nu_\psi(\epsilon)$ are the exponents 
for the chiral theory O(2)$\otimes$O($n_2$) and for the XY model, respectively.
At five loops we obtain 
\begin{equation}
n_2(\epsilon)= 23.1513-28.2072 \epsilon + 18.8689 \epsilon^2 - 
               27.1434 \epsilon^3  + 74.2373 \epsilon^4 + 
              O(\epsilon^5)\,.
\end{equation}
A Pad\'e-Borel analysis of this series gives $N_2(1)\approx 7$, 4, 7,
at five, four, and three loops.  The estimates nicely oscillate and
allow us to predict $4 \lesssim N_2(1) \lesssim 7$.  We also analyzed
the inverse series $1/n_2(\epsilon)$.  A Pad\'e-Borel analysis of the
series is possible only at five loops; at four loops all Pad\'e
approximants are defective. In this case we obtain $N_2(1)\approx 8$.
These results are fully compatible with the conclusions of
Sec.~\ref{sec3.3}.

\subsection{Renormalization-group
flow in the $3d$-$\overline{\rm MS}$ scheme for $N=3$}
\label{sec4.2}

In order to investigate the RG flow in the complete space of the
quartic parameters and check for the existence of other stable FPs, we
compute the five-loop series of the $\beta$ functions associated with
the five quartic couplings in the $\overline{\rm MS}$ scheme for the
general theory (\ref{LGWH}).  The diagrams contributing to this
calculation are a few hundreds.  We handled them with a symbolic
manipulation program, which generated the diagrams and computed the
symmetry and group factors of each of them. We used the results of
Ref.~\cite{KS-01}, where the primitive divergent parts of all
integrals appearing in our computation are reported.  The numerical
analysis of the series is very complex and long, so that we only study
the physical case $N=3$.
 
To find the stable FPs we follow the RG flow generated by the resummed
$\beta$ functions. For the resummation we use the Pad\'e-Borel method,
since the large-order behavior of the series, needed to perform the
conformal mapping, is not known. We use several different Pad\'e
approximants, the [4/1] and the [3/2] approximants at five loops, and
the [3/1] and the [2/2] at four loops.  We consider several initial
conditions close to the Gaussian FP, that allow us to explore a large
region in the space of the five renormalized couplings. For most of
the employed approximants we found runaway trajectories, up to where
the resummation is effective.  There are however two notable
exceptions. First, if we use the [4/1] Pad\'e approximant, we find
that the RG flow ends at the chiral O(2)$\otimes$O(3) FP. This is in
agreement with the conclusions of Sec.~\ref{sec3.2}: this FP may be
the stable FP of the model.  Second, at four loops, by using the [3/1]
Pad\'e approximant, one observes a stable FP with $v > 0$.  However,
there is no indication for such a FP at five loops, even as an
unstable FP.

In conclusion, the Pad\'e-Borel analysis of the RG flow does not
provide any evidence for a stable FP beside the chiral
O(2)$\otimes$O(3) FP. Thus, the multicritical transition is either
controlled by this FP or is of first order.

We also considered the multicritical model with $v=0$.  The
nonperturbative analysis of Sec.~\ref{sec3.3} indicates that the
decoupled collinear FP is stable for $N=3$. We also analyzed the full
model, checking whether other FPs are present. The five-loop analysis
did not find additional FPs, so that the multicritical behavior should
be controlled by the decoupled collinear FP (of course, when the
transition is continuous).

\section{Conclusions} \label{sec5}

\begin{figure}
\centerline{\epsfig{file=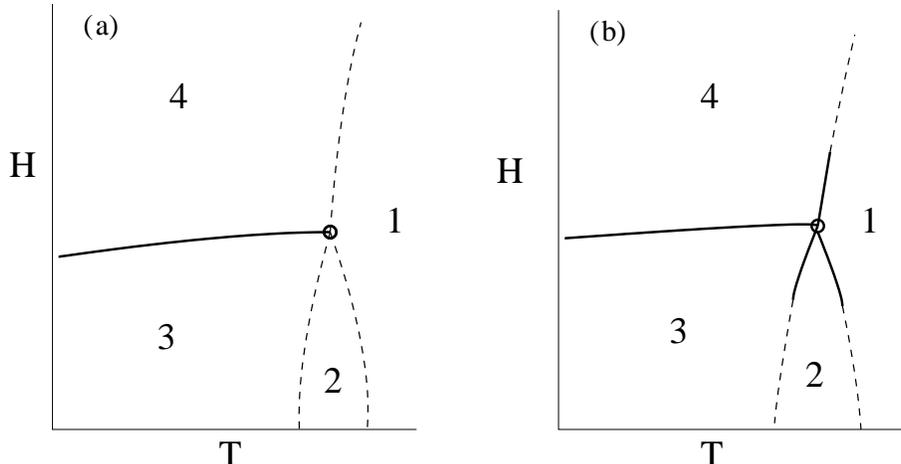,width=12truecm}}
\caption{Possible phase diagrams for easy-axis STAs in a magnetic 
field: (a) the multicritical transition is continuous; 
(b) the multicritical transition is of first order.
Continuous thick lines represent first-order transitions,
while the dashed lines correspond to second-order transitions.}
\label{fig-phd-pred}
\end{figure}

In this paper we study a general model with symmetry
O(2)$\otimes$[${\mathbb Z}_2\oplus$O($N-1$)] for $N\ge 3$, focusing on
the nature of the multicritical point. For $N\gtrsim 6$, the theory
presents a stable FP that is expected to control the multicritical
behavior: the decoupled FP where the fields $\phi$ and $\psi$ become
separately critical.  Their critical fluctuations are controlled
respectively by the chiral O(2)$\otimes$O($N-1$) FP and by the XY
FP. For $3 < N \lesssim 6$ neither the decoupled FP nor the chiral
O(2)$\otimes$O($N$) FP are stable. However, we have not performed a
thourough analysis of the RG flow, so that we cannot exclude that
there exists a nontrivial stable FP that does not belong to any
submodel we have investigated.  A complete analysis has been performed
for the physically interesting case $N=3$. In this case, we find that
the only possible stable FP is the chiral O(2)$\otimes$O(3) FP.  The
perturbative analysis in two different perturbative schemes is unable
to draw a definite conclusion on the stability of this FP.  The
possible phase diagrams are reported in Fig.~\ref{fig-phd-pred}.  In
phase diagram (a) the transition at the multicritical point is
continuous and controlled by the O(2)$\otimes$O(3) FP, thus showing
the symmetry enlargement originally put forward in
Ref.~\cite{KCP-90}; this scenario requires the stability of the
O(2)$\otimes$O(3) FP and that the system is within its attraction
domain.  
In the other possible cases, i.e. if the O(2)$\otimes$O(3) FP
is unstable or if the system is outside its attraction domain, we should
observe phase diagram (b), in which the 
multicritical transition is of first order; first-order transitions are
also expected along the lines separating phases 1-4, 1-2, and 1-3,
close to the multicritical point.

It is interesting to compare this scenario with the experimental
results \cite{CP-97}.  The behavior observed in experiments at the
multicritical point in always compatible with a second-order
transition.  Moreover, the experimental
estimates for the critical exponents (see the results for CsNiCl$_3$
and CsMnI$_3$ in Ref.~\cite{CP-97}) are reasonably close to the
theoretical results for the O(2)$\otimes$O(3) chiral universality
class. Therefore, the phase diagram presented in Fig.~\ref{fig-phd-pred}(a)
seems to be favored by experiments, even though the first-order
scenario is not necessarily ruled out.  If phase diagram (b) 
is the correct one, a possible explanation of the experiments is that
the first-order transition is rather weak, so that all experiments are
still probing a crossover region.  This interpretation may be
supported by the following reasoning. The experimental systems have a
small easy-axis anisotropy and therefore, see Sec.~\ref{sec2}, they are
approximately described by the effective theory with $w_0 \approx y_0
\approx z_0 \approx 0$.  Thus, the RG flow starts very close to the
O(2)$\otimes$O(3) FP, so that one expects strong crossover effects
controlled by the chiral O(2)$\otimes$O(3) theory. Moreover, the flow
out of the chiral O(2)$\otimes$O(3) FP should be very slow, since the
associated crossover exponent $\phi_{4,4}$ is very small,
$\phi_{4,4}\lesssim 0.1$.  The first-order nature of the multicritical
point also implies that along the lines separating phases 1-2 and 2-3,
one should observe first-order transitions, a tricritical point, and
then XY behavior. The presence of the tricritical point might explain
why at the transitions for $H=0$ one observes values of $\nu$ and
$\gamma$ that are significantly different from the XY estimates and
that are closer to the mean-field predictions (see the results for
CsMnI$_3$ reported in Ref.~\cite{KIANE-91}).  Finally, the presence of
the first-order transition might also explain the discrepancies
between the experimental estimates of the critical exponents along the
line 1-4 and the theoretical predictions for the O(2)$\otimes$O(2)
chiral universality class.  Indeed, as discussed in
Ref.~\cite{CPPV-04}, close to the first-order parameter region,
O(2)$\otimes$O(2) chiral systems show strong crossover effects, with
effective exponents that may significantly differ from their
asymptotic value.

\appendix

\section{Mean-field phase diagram} \label{AppA}

\subsection{Model with $v_0 = 0$} \label{AppA.1}

We begin by discussing the mean-field phase diagram for $v_0 = 0$,
which corresponds to the model with larger symmetry
[O(2)$\otimes$O($N-1$)]$\oplus$O(2). For the discussion it is useful
to introduce new couplings
\begin{eqnarray}
   g_1 &=& u_0 + w_0,  \nonumber \\
   g_2 &=& u_0 + y_0,  \nonumber \\
   g_3 &=& 2 u_0 + 2 w_0 - z_0.
\end{eqnarray}  
In terms of $g_1$, $g_2$, and $g_3$, the stability conditions for the 
quartic potential are particularly simple. We have $g_1 > 0$, 
$g_2 > 0$, $g_3 > 0$, and 
\begin{eqnarray}
u_0 > - \sqrt{g_2 g_3/2}  && \qquad \qquad 
\hbox{for } 0 < g_3 < 2 g_1,  \nonumber \\
u_0 > -\sqrt{g_1 g_2}     && \qquad \qquad 
\hbox{for } g_3 > 2 g_1.  \nonumber 
\end{eqnarray}     
In order to solve the mean-field equations it is useful to use the symmetry 
in order to obtain a simple parametrization of the fields. 
Using the O($N-1$) invariance for $\phi_{ai}$ and the O(2) invariance 
for $\psi$ we can write 
\begin{eqnarray}
  \phi_{a1} = (a,0,\ldots), \qquad
  \phi_{a2} = (b,c,0\ldots), \qquad
  \psi_{i} = (d,0). \
\end{eqnarray}   
Note that we have not used the additional O(2) symmetry transformations 
applied to the $\phi$ fields. If $a \not = 0$ and $b^2 + c^2 \not = 0$,
we can perform O(2)$\otimes$O($N-1$) transformations to set 
$a^2 = b^2 + c^2$. 

The mean-field equations are easily solved and we obtain six classes
of solutions (we report one representative for each class; other solutions 
in each class are obtained by applying the symmetry transformations):
\begin{enumerate}
\item[(a)]  $a=b=c=d=0$, with energy $H = 0$.
This is a minimum only if $r_1 > 0$ and $r_2 > 0$.
\item[(b)]  $\phi_1^2 = - 6 r_1/g_1$, $\phi_2 = 0$, $\psi^2 = 0$, with energy
       $H = - {3\over2} r_1^2/g_1$.
\item[(c)] $a^2 = c^2 = - 6 r_1/g_3$, $b=d=0$, 
with energy $H = - {3} r_1^2/g_3$.
\item[(d)] $\psi^2 = - 6 r_2/g_2$, $a=b=c=0$, 
with energy $H = - {3\over2} r_2^2/g_2 $.
\item[(e)] $a^2 = -6 (g_2 r_1 - r_2 u_0)/D$, $d^2 = -6 (g_1 r_2 - r_1 u_0)/D$,
$b^2 = c^2 = 0$,
with $H = - {3\over2} (g_2 r_1^2  + g_1 r_2^2  - 2 r_1 r_2 u_0)/D$,
with $D \equiv g_1 g_2 - u_0^2$.
\item[(f)] $a^2 = c^2 = -6 (g_2 r_1 - r_2 u_0)/D_2$,
$d^2 = -6 (g_3 r_2 - 2 r_1 u_0)/D_2$, with
$H = - {3\over2} (2 g_2 r_1^2  + g_3 r_2^2  - 4 r_1 r_2 u_0)/D_2$, with
$D_2 \equiv g_2 g_3 - 2 u_0^2$,
\end{enumerate} 
where $r_1 = r_\phi$ and $r_2 = r_\psi$. 
In order to determine the phase diagram, for each value of the couplings
and of the ratio $r_1/r_2$, we must determine which solution has the 
lowest energy. The results are the following: 

\begin{itemize}
\item[(1)] $0 < g_3 < 2 g_1$ and $u_0^2 < g_3 g_2/2$.
The critical point is tetracritical. For $r_1 > 0$ and $r_2 > 0$ the
system is paramagnetic. Then, proceeding anticlockwise in the plane
$(r_1,r_2)$ we find: a transition line for $r_1 = 0$ and $r_2 > 0$;
phase (c); a transition line for $r_2 = 2 u_0 r_1/g_3$ and $r_1 < 0$;
phase (f); a transition line for $r_2 = g_2 r_1/u_0$ and $r_2 < 0$;
phase (d); a transition line for $r_2 = 0$ and $r_1 > 0$.
All transitions are continuous. In the presence of fluctuations
transitions (a)/(d) and (c)/(f) belong to the XY universality class,
while transitions (a)/(c) and (d)/(f) are chiral transitions,
i.e. correspond to the symmetry breaking
O(2)$\otimes$O($N-1$)$\to$O(2)$\otimes$O($N-3$).    

\item[(2)] $0 < g_3 < 2 g_1$ and $u_0^2 > g_3 g_2/2$.
The critical point is bicritical. For $r_1 > 0$ and $r_2 > 0$ the
system is paramagnetic. Then, proceeding anticlockwise in the plane
$(r_1,r_2)$ we find: a transition line for $r_1 = 0$ and $r_2 > 0$;
phase (c); a transition line for $r_2 = r_1 \sqrt{2 g_2/g_3}$ and
$r_1 < 0$; phase (d); a transition line for $r_2 = 0$ and $r_1 > 0$.
Transitions (a)/(c) and (a)/(d) are continuous: 
transition (a)/(c) is 
a chiral one while 
transition (a)/(d) belongs to the XY universality class. 
Transition (c)/(d) is of first order.
 
\item[(3)] $g_3 > 2 g_1$ and $u_0^2 < g_1 g_2$.
The critical point is tetracritical. For $r_1 > 0$ and $r_2 > 0$ the
system is paramagnetic. Then, proceeding anticlockwise in the plane
$(r_1,r_2)$ we find: a transition line for $r_1 = 0$ and $r_2 > 0$;
phase (b); a transition line for $r_2 = u_0 r_1/g_1$ and $r_1 < 0$;
phase (e); a transition line for $r_2 = g_2 r_1/u_0$ and $r_2 < 0$;
phase (d); a transition line for $r_2 = 0$ and $r_1 > 0$.
All transitions are continuous. In the presence of fluctuations
transitions (a)/(d) and (b)/(e) belong to the XY universality class,
while transitions (a)/(b) and (d)/(e) are collinear transitions,
i.e. correspond to the symmetry breaking
O(2)$\otimes$O($N-1$)$\to {\mathbb Z}_2\otimes$O($N-2$).  

\item[(4)] $g_3 > 2 g_1$ and $u_0^2 > g_1 g_2$.
The critical point is bicritical. For $r_1 > 0$ and $r_2 > 0$ the
system is paramagnetic. Then, proceeding anticlockwise in the plane
$(r_1,r_2)$ we find: a transition line for $r_1 = 0$ and $r_2 > 0$;
phase (b); a transition line for $r_2 = r_1 \sqrt{g_2/g_1}$ and
$r_1 < 0$; phase (d); a transition line for $r_2 = 0$ and $r_1 > 0$.
Transitions (a)/(b) and (a)/(d) are continuous: 
transition (a)/(b) is 
a collinear one while 
transition (a)/(d) belongs to the XY universality class. 
Transition (b)/(d) is of first order.
 
\end{itemize}

\subsection{Model with $v_0 > 0$} \label{AppA.2}

We shall now focus on the case $v_0 \not=0$. We shall only consider the case 
$v_0 > 0$, since the case $v_0 < 0$ can be recovered by using 
the symmetry (\ref{sym1}). As before, we introduce new couplings 
\begin{eqnarray}
   g_1 &=& u_0 + w_0,  \nonumber \\
   g_2 &=& u_0 + y_0,  \nonumber \\
   g_3 &=& 2 u_0 + 2 w_0 - v_0 - z_0,
\end{eqnarray}  
that are all required to be positive by the stability of the quartic
potential and are invariant under the symmetry (\ref{sym1}). 
We also found additional necessary stability conditions.
First, we must have
\begin{eqnarray}
  u_0 &>& v_0 - \sqrt{g_1 g_2}, \nonumber \\
  u_0 &>&  - \sqrt{g_1 g_2}, 
\end{eqnarray}
where only the first condition is relevant for $v_0 > 0$. In order to 
write down the second condition let us define
\begin{eqnarray}
    R_1 &\equiv& {1\over 2} g_2 g_3 - {1\over 4} {g_3 v_0^2\over 2 g_1 - g_3},
\nonumber \\
    R_2 &\equiv& {v_0^2 \over 2( 2 g_1 - g_3) },
\nonumber \\
    R_3 &\equiv& (2 g_1 g_2 - g_2 g_3 - 2 g_1 u_0 + g_3 u_0 + g_1 v_0 - u_0 v_0)
\nonumber \\
    && \times
     (2 g_1 g_2 - g_2 g_3 - 2 g_1 u_0 + g_3 u_0 + g_1 v_0 - g_3 v_0 + u_0 v_0 -
      v_0^2 ),
\end{eqnarray}
and the domain $\Omega$ in the coupling space
\begin{equation}
\Omega = \{ (u_0,v_0,g_1,g_2,g_3) : \, 
        g_3 < 2 g_1, \,  g_2 > R_2, \, R_3 > 0\}.
\end{equation}
If $(u_0,v_0,g_1,g_2,g_3)\in \Omega$, then the couplings must satisfy
\begin{eqnarray}
    &&  u_0 > {v_0\over2} - \sqrt{R_1},
\end{eqnarray}
We have not been able to prove that these conditions are sufficient for the 
stability of the quartic potential. However, since they go over to the 
stability conditions for $v_0 = 0$ (in this case we have proved they are 
sufficient), we believe they are enough for the stability of the quartic 
potential.

We parametrize the $\phi_{ai}$ fields as before, while for $\psi_i$ we must
keep both components, i.e. we set $\psi_i = (d,e)$, because of the 
reduced symmetry of the model. Setting as before 
$r_1 = r_\phi$ and $r_2 = r_\psi$, the mean-field equations are
\begin{eqnarray}
&& a r_1 + {a\over 6} \left[ u_0 (\phi^2 + \psi^2) - v_0 (c^2 + e^2)
       + w_0 \phi^2 - z_0 c^2\right] + {1\over6} v_0 bde = 0,
\label{eqmf-1} \\
&& b r_1 + {b\over 6} \left[ u_0 (\phi^2 + \psi^2) - v_0 d^2
       + w_0 \phi^2\right] + {1\over6} v_0 ade = 0,
\label{eqmf-2} \\
&& c r_1 + {c\over 6} \left[ u_0 (\phi^2 + \psi^2) - v_0 (a^2 + d^2)
       + w_0 \phi^2 - z_0 a^2 \right] = 0,
\label{eqmf-3} \\
&& d r_2 + {d\over 6} \left[ u_0 (\phi^2 + \psi^2) - v_0 (b^2 + c^2)
       + y_0 \psi^2 \right] + {1\over 6} v_0 abe = 0,
\label{eqmf-4} \\
&& e r_2 + {e\over 6} \left[ u_0 (\phi^2 + \psi^2) - v_0 a^2
       + y_0 \psi^2 \right] + {1\over 6} v_0 abd = 0.
\label{eqmf-5} 
\end{eqnarray} 
The solutions are:
\begin{enumerate}
\item[(a)]  $a=b=c=d=e=0$, with energy $H = 0$.
\item[(b)]  $\phi_1^2 = - 6 r_1/g_1$, $\phi_2 = 0$, $\psi^2 = 0$, with energy
       $H = - {3\over2} r_1^2/g_1$.
\item[(c)] $a^2 = c^2 = - 6 r_1/g_3$, $b=d=e=0$,
with energy $H = - {3} r_1^2/g_3$.
\item[(d)] $\psi^2 = - 6 r_2/g_2$, $a=b=c=0$, 
with energy $H = - {3\over2} r_2^2/g_2 $.
\item[(e1)] $b^2 = -6 (g_2 r_1 - r_2 u_0)/D$, $e^2 = -6 (g_1 r_2 - r_1 u_0)/D$,
$a^2 = c^2 = d^2 = 0$,
with $H = - {3\over2} (g_2 r_1^2  + g_1 r_2^2  - 2 r_1 r_2 u_0)/D$,
with $D \equiv g_1 g_2 - u_0^2$.
\item[(e2)] $b^2 = -6 (g_2 r_1 - r_2 (u_0 - v_0))/D_3$,
            $d^2 = -6 (g_1 r_2 - r_1 (u_0 - v_0))/D_3$,
$a^2 = c^2 = e^2 = 0$,
with $H = - {3\over2} (g_2 r_1^2  + g_1 r_2^2  - 2 r_1 r_2 (u_0-v_0))/D_3$,
with $D_3 \equiv g_1 g_2 - (u_0-v_0)^2$.
\item[(f)] This solution is too long to be reported. Both vectors are 
nonvanishing, and we can take $a^2 = b^2 + c^2$, $b^2 = O(v_0^2)$.
Moreover, $d^2 = e^2$ and $de/(ab) = (g_3 -  2 g_1)/v_0$.
\end{enumerate}      
The solutions follow the labelling used for $v_0 = 0$. Note that for 
$v_0 > 0$ there are two different solutions corresponding the 
solution (e) found before. The derivation of these solutions is 
straightforward, except for case (f). To derive (f), assume that 
$a$, $b$, $c$, $d$, and $e$ are nonvanishing. Then, Eqs.~(\ref{eqmf-2})
and (\ref{eqmf-3}) imply 
\begin{equation}
d e =  a b {g_3 - 2 g_1 \over v_0}.
\label{constr}
\end{equation}
Substitute this relation in Eqs.~(\ref{eqmf-1}) 
and (\ref{eqmf-2}). Analogously, one can use this relation 
to express $ab$ in terms of $de$ in Eqs.~(\ref{eqmf-4}) and (\ref{eqmf-5}).
The five mean-field equations become linear in $a^2$, $b^2$,
$c^2$, $d^2$, and $e^2$. However, only three
of them are independent. To completely solve the problem, we consider 
Eqs.~(\ref{eqmf-1}), (\ref{eqmf-2}), and (\ref{eqmf-4}) that are independent, 
the equation $a^2 = b^2 + c^2$ that fixes the O(2) invariance, and the 
relation between $d^2 e^2$ and $a^2 b^2$ that follows from Eq.~(\ref{constr}). 
This system of equations has a unique solution for $a^2$, $b^2$, $c^2$,
$d^2$, and $e^2$.

The analysis of the phase diagram of this model is extremely complex, 
mainly due to the cumbersome expressions for solution (f). We have used 
analytic and numerical methods to sort out the different possibilities. 
We find: 

\begin{itemize} 

\item[(a)] We have a pentacritical point for $2 g_1 - g_3 > 0$, $g_2 > R_2$,
and $v_0/2 -\sqrt{R_1} < u_0 < v_0/2 + \sqrt{R_1}$. 
For $r_1 > 0$ and $r_2 > 0$ the
system is paramagnetic. Then, proceeding anticlockwise in the plane
$(r_1,r_2)$ we find: a transition line for $r_1 = 0$ and $r_2 > 0$;
phase (c); a transition line for $r_1 = g_3 r_2/(2 u_0 - v_0)$ and $r_1 < 0$;
phase (f); a transition line;
phase (e2); a transition line for $r_1 = (u_0 - v_0) r_2/g_2$ and $r_2 < 0$;
phase (d); a transition line for $r_2 = 0$ and $r_1 > 0$.
All transitions are continuous except that between phases 
(f) and (e2) that is of first order. 
In the presence of fluctuations
transitions (a)/(d) and (c)/(f) belong to the XY universality class,
transition (d)/(e2) belongs to the O($N-1$) vector universality class, 
while transition (a)/(c) is a chiral transition,
i.e. it corresponds to the symmetry breaking
O(2)$\otimes$O($N-1$)$\to$O(2)$\otimes$O($N-3$). For $v_0 \to 0$ the width of 
phase (e2) goes to zero and we obtain case (1) considered in the 
previous Section.

\item[(b)] We have a tetracritical point in three cases: 
$2 g_1 - g_3 > 0$, $g_2 > R_2$, 
$u_0 < v_0/2 - \sqrt{R_1}$ up to the stability boundary;
$2 g_1 - g_3 > 0$, $g_2 > R_2$,     
$v_0/2 + \sqrt{R_1} < u_0 < v_0 + \sqrt{g_2 g_3/2}$;
$2 g_1 - g_3 > 0$, $ 0 < g_2 < R_2$, 
$u_0 < v_0 + \sqrt{g_2 g_3/2}$ up to the stability boundary.
For $r_1 > 0$ and $r_2 > 0$ the
system is paramagnetic. Then, proceeding anticlockwise in the plane
$(r_1,r_2)$ we find: a transition line for $r_1 = 0$ and $r_2 > 0$;
phase (c); a transition line with $r_1 < 0$;
phase (e2); a transition line with $r_1 = (u_0-v_0) r_2/g_2$ and $r_2 < 0$;
phase (d); a transition line for $r_2 = 0$ and $r_1 > 0$.
Transitions (a)/(c), (e2)/(d), (d)/(a) are continuous, while transition
(c)/(e2) is of first order.
In the presence of fluctuations
transition (a)/(d) belongs to the XY universality class,
transition (d)/(e2) belongs to the O($N-1$) vector universality class, 
while transition (a)/(c) is a chiral transition,
i.e. it corresponds to the symmetry breaking
O(2)$\otimes$O($N-1$)$\to$O(2)$\otimes$O($N-3$). 
Such a case does not exist for $v_0 = 0$.

\item[(c)] We have a bicritical point for
$2 g_1 - g_3 > 0$ and
$u_0 > v_0 + \sqrt{g_2 g_3/2}$.
For $r_1 > 0$ and $r_2 > 0$ the
system is paramagnetic. Then, proceeding anticlockwise in the plane
$(r_1,r_2)$ we find: a transition line for $r_1 = 0$ and $r_2 > 0$;
phase (c); a transition line for $r_1 = r_2 (g_3/2 g_2)^{1/2}$;
phase (d); a transition line for $r_2 = 0$ and $r_1 > 0$.
Transition (c)/(d) is of first order, transition (a)/(c) is a chiral 
transition, while 
transition (a)/(d) belongs to the XY universality class.

\item[(d)] We have a tetracritical point for
$2 g_1 - g_3 < 0$ and 
$v_0 - \sqrt{g_1 g_2} < u_0 < v_0 + \sqrt{g_1 g_2}$.
For $r_1 > 0$ and $r_2 > 0$ the
system is paramagnetic. Then, proceeding anticlockwise in the plane
$(r_1,r_2)$ we find: a transition line for $r_1 = 0$ and $r_2 > 0$;
phase (b); a transition line for $r_1 = g_1 r_2/(u_0 - v_0)$ and $r_1 < 0$;
phase (e2); $r_1 = (u_0 - v_0) r_2/g_2$ and $r_2 < 0$;
phase (d); a transition line for $r_2 = 0$ and $r_1 > 0$. 
All transitions are continuous. Transition (a)/(b) 
is a collinear transition with $v_0 < 0$, i.e. it corresponds to the 
symmetry breaking O(2)$\otimes$O($N-1$)$\to {\mathbb Z}_2 \otimes$ O($N-2$), 
transitions (b)/(e2) and (a)/(d) are XY transitions, 
while transition (d)/(e2) belongs to the O($N-1$) vector universality class.

\item[(e)] We have a bicritical point for $2 g_1 - g_3 < 0$ and
$u_0 > v_0 + \sqrt{g_1 g_2}$.      
For $r_1 > 0$ and $r_2 > 0$ the
system is paramagnetic. Then, proceeding anticlockwise in the plane
$(r_1,r_2)$ we find: a transition line for $r_1 = 0$ and $r_2 > 0$;
phase (b); a transition line for $r_1 = r_2 (g_1/g_2)^{1/2}$; 
phase (d); a transition line for $r_2 = 0$ and $r_1 > 0$.
Transition (a)/(b) 
is a collinear transition with $v_0 < 0$, 
transition (b)/(d) is of first order, and 
transition (a)/(d) belongs to the XY universality class.

\end{itemize}

\section{Classification of the perturbations of an O($M$)$\otimes$O($N$) 
theory} \label{AppB}

We consider an O($M$)$\otimes$O($N$)-symmetric theory with  
$M,N\ge 2$ and the Hamiltonian
\begin{eqnarray}
{\cal H}_{MN} &=& \int d^d x 
 \Bigl\{ {1\over2}
\sum_{ai} \Bigl[ (\partial_\mu \Phi_{ai})^2 + r \Phi_{ai}^2 \Bigr] \nonumber \\
&&
+ {1\over 4!}u_0 \Bigl( \sum_{ai} \Phi_{ai}^2\Bigr)^2
+ {1\over 4!}  v_0
\sum_{a,b} \Bigl[ ( \sum_i \Phi_{ai} \Phi_{bi})^2 - (\sum_i \Phi_{a}^2)(\sum_i\Phi_{bi})^2\Bigr]
             \Bigr\},
\label{LGWHch}
\end{eqnarray}
where $\Phi_{ai}$ ($a=1,...N$ and $i=1,...M$) are 
$N\times M$ matrix variables. We wish now to 
classify the quadratic and quartic operators that 
break the  O($N$) symmetry but preserve the O($M$) symmetry. 
The classification of all the quadratic operators that break both 
symmetries is presented in App.~\ref{AppC}.

At the quadratic level, there is only one operator, the spin-2 operator
\begin{equation}
{\cal O}^{(2,2)}_{ab} = \Phi_a \cdot \Phi_b - {1\over N} \delta_{ab} \Phi^2,
\end{equation}
where (as in all this Section) the scalar product is in the O($M$) space 
and $\Phi^2 = \sum_a \Phi_a\cdot \Phi_a$.

At the quartic level, there are three operators, one spin-4 operator,
two spin-2 operators and one operator associated with a 
nontrivial O($N$) representation. The spin-4 operator is given by
\begin{eqnarray}
{\cal O}^{(4,4)}_{abcd} &=&
   (\Phi_a \cdot \Phi_b) (\Phi_c\cdot \Phi_d) + \hbox{2 perm.} 
\nonumber \\
   && - {1\over N +4} \Phi^2 (\delta_{ab} (\Phi_c\cdot \Phi_d) + \hbox{5 perm.})
\nonumber \\
   && - {2\over N +4} (\delta_{ab} \sum_e (\Phi_e\cdot \Phi_c)
       (\Phi_e\cdot \Phi_d) + \hbox{5 perm.})
\nonumber \\
   && + {1\over (N+2)(N+4)} (\Phi^2)^2 
        (\delta_{ab} \delta_{cd} + \hbox{2 perm.})
\nonumber \\
   && + {2\over (N+2)(N+4)} [\sum_{ef} (\Phi_e\cdot\Phi_f)^2]
      (\delta_{ab} \delta_{cd} + \hbox{2 perm.}),
\end{eqnarray}
where the permutations are such to make ${\cal O}^{(4,4)}_{abcd}$
symmetric and traceless. The spin-2 operators are given by
\begin{eqnarray}
{\cal O}^{(4,2,1)}_{ab} &=& \Phi^2 {\cal O}^{(2,2)}_{ab} ,
\nonumber \\
{\cal O}^{(4,2,2)}_{ab} &=& \sum_e (\Phi_a \cdot \Phi_e) (\Phi_b \cdot \Phi_e) 
       - {1\over N} \delta_{ab} [\sum_{ef} (\Phi_e\cdot\Phi_f)^2].
\end{eqnarray}
They are always independent except for $N=2$. In this case 
${\cal O}^{(4,2,1)}_{ab} = {\cal O}^{(4,2,2)}_{ab}$. 

The remaining operator is 
\begin{eqnarray}
{\cal O}^{(4,r)}_{abcd} &=& (\Phi_a \cdot \Phi_c) (\Phi_b \cdot \Phi_d) - 
         (\Phi_a \cdot \Phi_d) (\Phi_b \cdot \Phi_c) 
\nonumber \\
       && - {1\over N(N-1)} (\delta_{ac}\delta_{bd} - \delta_{ad}\delta_{bc})
          [(\Phi^2)^2 - \sum_{ef} (\Phi_e\cdot\Phi_f)^2]
\nonumber \\
  && - {\delta_{ac}\over N - 2} 
     ({\cal O}^{(4,2,1)}_{bd} - {\cal O}^{(4,2,2)}_{bd})
     - {\delta^{bd}\over N - 2} 
     ({\cal O}^{(4,2,1)}_{ac} - {\cal O}^{(4,2,2)}_{ac})
\nonumber \\
  && + {\delta^{ad}\over N - 2} 
     ({\cal O}^{(4,2,1)}_{bc} - {\cal O}^{(4,2,2)}_{bc})
     + {\delta^{bc}\over N - 2} 
     ({\cal O}^{(4,2,1)}_{ad} - {\cal O}^{(4,2,2)}_{ad})\; .
\end{eqnarray}
Such a quantity satisfies the properties
${\cal O}^{(4,r)}_{abcd} = - {\cal O}^{(4,r)}_{bacd}$, 
${\cal O}^{(4,r)}_{abcd} = - {\cal O}^{(4,r)}_{abdc}$,
${\cal O}^{(4,r)}_{abcd} = {\cal O}^{(4,r)}_{cdab}$, 
$\sum_{c=1}^N {\cal O}^{(4,r)}_{cacb} = 0$.
It belongs to a nontrivial representation of the O($N$) group, the one 
associated with the Young tableau that has the shape of a 2$\times$2 box.
(Note that in our terminology we call spin-$k$ representation the 
representation associated with the Young tableau that has the shape of 
a row of length $k$).
This representation occurs only for $N\ge 4$. Indeed, this 
operator is not defined for $N=2$; for $N=3$ we have 
$\sum_{abcd} \epsilon^{eab}\epsilon^{fcd} {\cal O}^{(4,r)}_{abcd} = 0$,
which shows that the operator does not occur.
The operator ${\cal O}^{(4,r)}_{abcd}$ does not mix with any other (quadratic 
or quartic) operator.

These operators control the symmetry breaking 
\begin{equation}
    O(M)\otimes O(N) \to O(M)\otimes [O(k) \oplus O(N-k)]\; .
\end{equation}
The corresponding multicritical Hamiltonian is 
\begin{equation}
{\cal H}_{N,k} = {\cal H}_{MN} + m_2 V^{(2,2)} + f_1 V^{(4,4)} + 
      f_2 V^{(4,2,1)} + f_3 V^{(4,2,2)} + f_4 V^{(4,r)},
\end{equation}
where 
$ V^{(2,2)} =\sum_{a=1}^k {\cal O}^{(2,2)}_{aa}$,
$V^{(4,4)} = \sum_{a=1}^k \sum_{b=k+1,n} {\cal O}^{(4,4)}_{aabb}$,
$V^{(4,2,1)} = \sum_{a=1}^k {\cal O}^{(4,2,1)}_{aa}$,
$V^{(4,2,2)} = \sum_{a=1}^k {\cal O}^{(4,2,2)}_{aa}$, 
$V^{(4,r)} = \sum_{ab=1}^k {\cal O}^{(4,r)}_{abab}$. 
For $k = 1$ or $k = N-1$, $V^{(4,r)}$ vanishes. Indeed, if $k=1$
$ V^{(4,r)} = {\cal O}^{(4,r)}_{1111}$ that vanishes because of 
the antisymmetry of the indices. For $k = N-1$, using the fact that 
${\cal O}^{(4,r)}_{abcd}$ is traceless, we have 
$V^{(4,r)} = - \sum_{a=1}^k {\cal O}^{(4,r)}_{aNaN} = 
         {\cal O}^{(4,r)}_{NNNN} = 0$.

Explicitly, if $\Phi_{am} \to (\phi_{Am}, \psi_{\alpha m})$ with 
$A=1,\ldots k$, $\alpha =1,\ldots N-k$, $m=1,\ldots M$, we have 
\begin{eqnarray}
V^{(2,2)} &=& {N-k\over N}\phi^2 - {k\over N} \psi^2, \\
\nonumber \\
  V^{(4,4)} &=& {1\over (N+2)(N+4)} \left\{-(N-k)(N-k+2) 
                [(\phi^2)^2 + 2 \sum_{AB} (\phi_A\cdot\phi_B)^2] \right.
\nonumber  \\
&& \left. - 
      k(k+2) [(\psi^2)^2 + 2 \sum_{\alpha\beta} (\psi_\alpha\cdot\psi_\beta)^2] 
                + 2 (k+2)(N-k+2) 
         [\phi^2 \psi^2 + 2 \sum_{A\alpha} (\phi_A\cdot\psi_\alpha)^2]\right\},
\nonumber \\
  V^{(4,2,1)} &=& (\phi^2 + \psi^2) V^{(2,2)} ,
\nonumber \\
  V^{(4,2,2)} &=& {N-k\over N}\sum_{AB} (\phi_A\cdot\phi_B)^2 + 
       {N-2 k\over N}\sum_{A\alpha} (\phi_A\cdot\psi_\alpha)^2 - 
       {k\over N} \sum_{\alpha\beta} (\psi_\alpha\cdot\psi_\beta)^2,
\nonumber \\
  V^{(4,r)} &=& - {1\over (N-1)(N-2)} \left\{-(N-k)(N-k-1) 
                [(\phi^2)^2 - \sum_{AB} (\phi_A\cdot\phi_B)^2] \right.
\nonumber  \\
&& \left. - 
     k(k-1) [(\psi^2)^2 - \sum_{\alpha\beta} (\psi_\alpha\cdot\psi_\beta)^2] 
                + 2 (k-1)(N-k-1) 
       [\phi^2 \psi^2 - \sum_{A\alpha} (\phi_A\cdot\psi_\alpha)^2]\right\}\; .
\end{eqnarray}
It is trivial to check that $V^{(4,r)}$ vanishes for $k = 1$ or 
$k = N-1$. 

In conclusion, at the quartic level: 
for $N=2$ there are two breaking operators,
$V^{(4,2,1)}$ and $V^{(4,4)}$; for $N=3$ and for $N\ge 4$, $k=1$ or $k = N-1$,
there are three breaking operators, $V^{(4,2,1)}$, $V^{(4,2,2)}$, and 
$V^{(4,4)}$; for $N\ge 4$ and $2\le k \le N-2$, all four operators should be 
considered. 

These operators are also relevant in other cases. 
If we consider the breaking 
\begin{equation}
    O(M)\otimes O(N) \to O(M) \otimes C_N,
\end{equation}
where $C_N$ is the cubic group in an $N$-dimensional space, we obtain the 
Hamiltonian
\begin{equation} 
   {\cal H}_{MN} + f \sum_{a=1}^N {\cal O}^{(4,4)}_{aaaa}\; .
\end{equation} 
For $N=4$ we may consider the breaking
\begin{equation}
    O(M)\otimes O(4) \to O(M) \otimes SO(4).
\end{equation}
The corresponding Hamiltonian is
\begin{equation}
   {\cal H}_{MN} + 
    f \sum_{abcd=1}^N \epsilon^{abcd} {\cal O}^{(4,r)}_{abcd} .
\end{equation}
Note, finally, that reduction to smaller symmetry groups does not require
the consideration of additional operators, although there may be 
additional terms in the Hamiltonian. For instance, the breaking 
\begin{equation}
    O(M)\otimes O(N) \to O(M) \otimes (C_k \oplus C_{N-k})
\end{equation}
is obtained by considering 
\begin{equation}
  {\cal H}_{N,k} + f_5 \sum_{a=1}^k {\cal O}^{(4,4)}_{aaaa} + 
            f_6 \sum_{b=k+1}^N {\cal O}^{(4,4)}_{bbbb}.
\end{equation}
The operators proportional to $f_1$, $f_5$, and $f_6$ are of course degenerate
at the O($M$)$\otimes$O($N$) FP.

\section{RG dimensions of the quadratic perturbations at the 
O(2)$\otimes$O($N$) fixed points}
\label{AppC}

In this appendix we consider the three-dimensional 
O(2)$\otimes$O($N$) invariant theory,
cfr. Eq.~(\ref{LGWHch}) with $M=2$, and compute 
the RG dimensions of all quadratic operators
breaking the O(2)$\otimes$O($N$) symmetry at the 
O(2)$\otimes$O($N$) FPs.

For generic $M$,
the quadratic operators breaking the O($M$)$\otimes$O($N$) symmetry
are explicitly given by
\begin{eqnarray}
O^{(1)}_{aibj} &=& \Phi_{ai}\Phi_{bj} -\Phi_{aj}\Phi_{bi},\\
O^{(2)}_{aibj} &=& \case{1}{2} 
    \left( \Phi_{ai}\Phi_{bj}+\Phi_{aj}\Phi_{bi}\right)
- \case{1}{N} \delta_{ab} \sum_c \Phi_{ci}\Phi_{cj} - 
  \case{1}{M} \delta_{ij} \sum_k\Phi_{ak}\Phi_{bk} 
+ \case{1}{MN} \delta_{ab} \delta_{ij} \sum_{ck} \Phi_{ck}\Phi_{ck}, \\
O^{(3)}_{ab} &=& \sum_{k} \Phi_{ak}\Phi_{bk} 
- \case{1}{N} \delta_{ab} \sum_{ck} \Phi_{ck}\Phi_{ck}, \\
O^{(4)}_{ij} &=& \sum_{c}\Phi_{ci}\Phi_{cj} 
- \case{1}{M} \delta_{ij} \sum_{ck} \Phi_{ck}\Phi_{ck},
\label{o4}
\end{eqnarray}
where $\Phi_{ai}$ is a real field with $a=1,\ldots,N$ and $i=1,\ldots,M$. 
These operators have a simple group-theory intepretation, that allows us 
to check that the list is exhaustive.
The operator $O^{(1)}_{aibj}$ transforms as a spin-1 operator under both 
O($N$) and O($M$), $O^{(2)}_{aibj}$ transforms as a spin-2 operator
under both groups, while $O^{(3)}_{ab}$ and $O^{(4)}_{ij}$ 
transform as a scalar under one group and as a spin-2 operator under the 
second group. For $M=2$ these operators correspond to those reported in 
Ref.~\cite{Kawamura-88}.

For $M=2$, we computed the RG dimensions $y_i$ of the
above-reported quadratic operators to six loops in the MZM scheme and to five
loops in the $3d$-$\overline{\rm MS}$ scheme.  We used a symbolic program to
generate diagrams and group factors and the compilations of Feynman integrals
of Refs.~\cite{NMB-77,KS-01}.
We do not report the series that are available on request.
The results of the analyses,
using the resummation methods outlined in
Refs.~\cite{CPV-00,PRV-01,CPPV-04}, are reported in Table~\ref{tabquadr}
for several values of $N$.  
In the case $N=6$ we do not report results for the chiral FP in the MZM scheme,
since in this scheme there is little evidence for the 
existence of a FP.
The RG dimensions of the operator $O^{(1)}$ related to the chiral degrees
of freedom at the chiral FP have already been computed by exploiting the same FT
methods in Refs.~\cite{PRV-02,CPPV-04}; we report them here for the
sake of completeness.  We also mention that these exponents have been computed
to order $1/N$ in Ref.~\cite{Kawamura-88}, while 
Ref.~\cite{Gracey-02} reports a $1/N^2$ calculation of the RG
dimension of $O^{(1)}$.

For $N=2$ the RG dimensions at the collinear FP can be related
to the RG dimensions of operators in the XY model. Indeed,
the O(2)$\otimes$O(2) collinear FP is equivalent to an XY FP.
The mapping is the following.
One defines two fields $a_i$ and $b_i$, $i=1,2$, and considers
\cite{Kawamura-88}
\begin{eqnarray}
&& \phi_{11} = (a_1 - b_2)/\sqrt{2},
\nonumber \\
&& \phi_{22} = (a_1 + b_2)/\sqrt{2},
\nonumber \\
&& \phi_{12} = (b_1 - a_2)/\sqrt{2},
\nonumber \\
&& \phi_{21} = (b_1 + a_2)/\sqrt{2}.
\end{eqnarray}
At the collinear FP, fields $a$ and $b$ represent two independent XY fields.
Using this mapping it is easy to show that:
$O^{(1)}_{aibj} \sim a^2 + b^2$;
$O^{(2)}_{aibj}$ is the sum of $a_1^2 - a_2^2$, $b_1^2 - b_2^2$,
$a_1 a_2$, and $b_1 b_2$; $O^{(3)}_{ab}$ (or $O^{(4)}_{ij}$) is the sum of
terms of the form $a_i b_j$. Thus,
$y_1 = y_{t,{\rm XY}} = 1/\nu_{\rm XY}$,
$y_2 = y_{T,{\rm XY}}$, where $y_{T,{\rm XY}}$ 
is the RG dimension of the spin-2 quadratic operator
in the XY model, and
$y_3 = y_{4} = 2 y_{h,{\rm XY}} - 3$, where
$y_{h,{\rm XY}} = (5 - \eta_{\rm XY})/2$ is the RG dimension of the
field in the XY model. Note also that the scalar operator $\phi^2$
becomes $a^2 + b^2$, as obviously expected.
Estimates of $\nu_{\rm XY}$, $\eta_{\rm XY}$, and $y_{T,{\rm XY}}$
can be found in Refs.~\cite{CHPRV-01,PV-rev,CPV-03,CPV-04nova}.

\begin{table}
\squeezetable
\caption{
RG dimensions of the quadratic operators breaking the symmetry
$O(2)\otimes O(N)$ at the chiral (ch) and collinear (cl) FPs.
We report the estimates obtained by analyzing 
the $\overline{\rm MS}$ (five loops) and MZM (six loops) expansions.
The errors include the spread of the considered approximants 
and the uncertainty on the location of the FP.
The results for $N=2$ at the collinear FP have been obtained by using the 
mapping with the XY model and the numerical results reported in 
Refs.~\protect\cite{CHPRV-01,CPV-03}.
}
\begin{tabular}{cllllllll}
\multicolumn{1}{c}{FP,$N$}& 
\multicolumn{2}{c}{$y_{1}$}&
\multicolumn{2}{c}{$y_{2}$}&
\multicolumn{2}{c}{$y_{3}$}&
\multicolumn{2}{c}{$y_{4}$}\\
\multicolumn{1}{c}{}& 
\multicolumn{1}{c}{$\overline{\rm MS}$}&
\multicolumn{1}{c}{MZM}&
\multicolumn{1}{c}{$\overline{\rm MS}$}&
\multicolumn{1}{c}{MZM}&
\multicolumn{1}{c}{$\overline{\rm MS}$}&
\multicolumn{1}{c}{MZM}&
\multicolumn{1}{c}{$\overline{\rm MS}$}&
\multicolumn{1}{c}{MZM}\\
\hline\hline
ch,2 & 2.37(15) & 2.54(12) & 2.00(15) & 2.07(7) & 1.34(15) & 
1.25(4) & 1.34(15) & 1.25(4) \\

ch,3 & 2.25(12) & 2.35(13) & 1.96(11) & 1.99(4) & 1.54(8) & 1.49(3)& 1.21(9) & 1.09(5) \\

ch,4  & 2.17(10) & 2.29(8) & 1.94(10) & 2.04(20) & 1.65(3) & 1.64(5) & 1.17(8) & 1.06(7) \\

ch,5  & 2.05(10) & 2.20(7) & 1.93(10) & 1.98(20) & 1.72(4) & 1.72(8) & 1.15(7) & 1.02(8) \\

ch,6  & 2.03(7) & & 1.90(10) & & 1.76(4) &  & 1.13(9) & \\ 

ch,8  & 2.02(2) & 2.03(4) & 1.92(2) & 1.93(4) & 1.81(2) & 1.79(1) & 1.13(8) & 1.13(4) \\


ch,16  & 2.001(5) & 2.00(1) & 1.948(5) & 1.95(1)  &  1.897(7) & 1.885(5) & 1.08(2) & 1.07(1)  \\

ch,$\infty$ & \multicolumn{2}{c}{2} & 
              \multicolumn{2}{c}{2} & 
              \multicolumn{2}{c}{2} & 
              \multicolumn{2}{c}{1} \\

cl,2 &        \multicolumn{2}{c}{$y_{t,\rm XY}=1.489(6)$} & 
              \multicolumn{2}{c}{$y_{T,{\rm XY}}=1.766(6)$} & 
              \multicolumn{2}{c}{$2y_{h,{\rm XY}}-3= 1.9620(8)$} & 
              \multicolumn{2}{c}{$2y_{h,{\rm XY}}-3= 1.9620(8)$} \\

cl,3  & 1.2(1) & 1.15(10) & 1.75(5) & 1.75(10) & 2.0(1) & 2.0(2) & 2.1(2) & 
        2.05(15) \\
cl,4  & 1.1(1) & 1.10(15) & 1.65(10) & 1.66(5) & 1.90(15) & 1.75(10) & 
        2.0(3) & 2.05(15) \\
\end{tabular}
\label{tabquadr}
\end{table}

\end{document}